\documentclass[a4paper,11pt,fleqn]{cas-sc}

\usepackage[authoryear]{natbib}
\usepackage{epsfig,color,amsmath}
 
\usepackage{wrapfig}
\usepackage{setspace}  
\usepackage{xcolor}
\usepackage{amssymb}
\usepackage{url}
\usepackage{mathtools}
\usepackage{enumitem}
\usepackage{multirow}
\usepackage{makecell}
\usepackage{stackrel}
\usepackage{booktabs}
\usepackage[flushleft]{threeparttable}
\usepackage{makecell}
\usepackage{subfig}
\usepackage{graphicx}

\usepackage[linesnumbered,lined,boxed,commentsnumbered,ruled,longend]{algorithm2e}


\setlist[itemize]{leftmargin=*}

\makeatother

\newcommand{\m}{\boldsymbol}
\allowdisplaybreaks[4]

\newcommand{\mbb}[1]{\mathbb{#1}}
\newcommand{\mr}[1]{\mathrm{#1}}

\usepackage{amsthm}

\usepackage[usestackEOL]{stackengine}
\stackMath
\usepackage[framemethod=TikZ]{mdframed}
\usepackage{bm}
\mdfdefinestyle{MyFrame}{%
	linecolor=black,
	outerlinewidth=1.5pt,
	roundcorner=1.5pt,
	innerrightmargin=5pt,
	innerleftmargin=5pt,}

\usepackage{tikz}
\usetikzlibrary{matrix,positioning,decorations.pathreplacing}

\usepackage[export]{adjustbox}
\usepackage{amsmath}
\usepackage{lipsum}
\usepackage{mathtools}
\usepackage{cuted}
\usepackage{physics}
\usepackage{enumerate}
\usepackage{lineno}
\usepackage{tabularray}
\usepackage{rotating}

\def\tsc#1{\csdef{#1}{\textsc{\lowercase{#1}}\xspace}}
\tsc{WGM}
\tsc{QE}

\begin{document}
\let\WriteBookmarks\relax
\def\floatpagepagefraction{1}
\def\textpagefraction{.001}

\shorttitle{{\tiny \singlespacing Real-time Regulation of Detention Ponds via Feedback Control: Balancing Flood Mitigation and Water Quality
}}   

\shortauthors{Gomes Jr. et. al}  


\title [mode = title]{Real-time Regulation of Detention Ponds via Feedback Control: Balancing Flood Mitigation and Water Quality
}  



\affiliation[1]{organization={University of São Paulo, Department of Hydraulic Engineering and Sanitation, São Carlos School of Engineering},
            addressline={Av. Trab. São Carlense, 400 - Centro}, 
            city={São Carlos},
            postcode={13566-590}, 
            state={São Paulo},
            country={Brazil}}
            
\affiliation[2]{organization={University of Texas at San Antonio, College of Engineering and Integrated Design, School of Civil \& Environmental Engineering and Construction Management},
            addressline={One UTSA Circle, BSE 1.310}, 
            city={San Antonio},
            postcode={78249}, 
            state={Texas},
            country={United States of America}}    

\affiliation[3]{organization={University of Arizona, Department of Hydrology and Atmospheric Sciences},
            addressline= {James E. Rogers Way, 316A}, 
            city={Tucson},
            postcode={85719}, 
            state={Arizona},
            country={United States of America}} 
            
\affiliation[4]{organization={Vanderbilt University, Department of Civil and Environmental Engineering},
            addressline= {Jacobs Hall, Office $\#$ 293, 24th Avenue South}, 
            city={Nashville},
            postcode={37235}, 
            state={Tennessee},
            country={United States of America}}

\author[1,2,3]{Marcus Nóbrega {Gomes Jr.}}[
        orcid=0000-0002-8250-8195X,
        ]
\cormark[1]


\ead{marcusnobrega.engcivil@gmail.com}

\ead[url]{www.engenheiroplanilheiro.com.br}

\credit{Conceptualization, Methodology, Software, Validation, Formal analysis, Investigation, Data Curation, Writing - Original Draft, Writing - Review \& Editing, Visualization, Resources}

\author[4]{Ahmad F. Taha}[
        orcid=0000-0003-0486-2794,
        ]
\ead{ahmad.taha@vanderbilt.edu}

\credit{Writing - Review \& Editing, Methdology}

\author[1]{Luis Miguel Castillo {Rápallo}}[
            orcid = 0000-0002-6241-7069,
        ]
        
\credit{Writing - Review \& Editing, Methdology, Investigation}

\ead{luis.castillo@usp.br }

\author[1]{Eduardo Mario Mendiondo}[
        orcid=0000-0003-2319-2773,
        ]
\ead{emm@sc.usp.br}

\credit{Writing - Review \& Editing, Funding acquisition, Resources}

\author[2]{Marcio Hofheinz Giacomoni}[
            orcid=0000-0001-7027-4128,
            ]
\credit{Writing - Review \& Editing, Funding acquisition, Visualization, Resources}
\ead{marcio.giacomoni@utsa.edu}

\cortext[1]{(corresponding author)}


\begin{abstract}
Detention ponds can mitigate flooding and improve water quality by allowing the settlement of pollutants. Typically, they are operated with fully open orifices and weirs (i.e., passive control). Active controls can improve the performance of these systems: orifices can be retrofitted with controlled valves, and spillways can have controllable gates.  The real-time optimal operation of its hydraulic devices can be achieved with techniques such as Model Predictive Control (MPC). A distributed quasi-2D hydrologic-hydrodynamic coupled with a reservoir flood routing model is developed and integrated with an MPC algorithm to estimate the operation of valves and movable gates in real-time. The control optimization problem is adapted to switch from a flood-related algorithm focusing on mitigating floods to a heuristic objective function that aims to increase the detention time when no inflow hydrographs are predicted. The case studies show the potential results of applying the methods developed in a catchment in Sao Paulo, Brazil. The performance of MPC compared to alternatives that do not change the operation over time with either fully or partially open valves and gates are tested. Comparisons with HEC-RAS 2D indicate volume and peak flow errors of approximately 1.4\%  and 0.91\% for the watershed module. Simulating two consecutive 10-year storms shows that the MPC strategy can achieve peak flow reductions of 79\%. In contrast, the passive scenario has nearly half of the performance (41\%). A 1-year continuous simulation results show that the passive scenario with 25\% of the valves opened can treat 12\% more runoff compared to the developed MPC approach, with an average detention time of approximately 6 hours. For the MPC approach, however, the average detention time is nearly 14 hours, indicating that both control techniques can treat similar volumes; however, the proxy water quality for the MPC approach is enhanced due to the longer detention times achieved. 
\end{abstract}




\maketitle


\section*{Graphical Abstract}

\includegraphics[scale = 0.60]{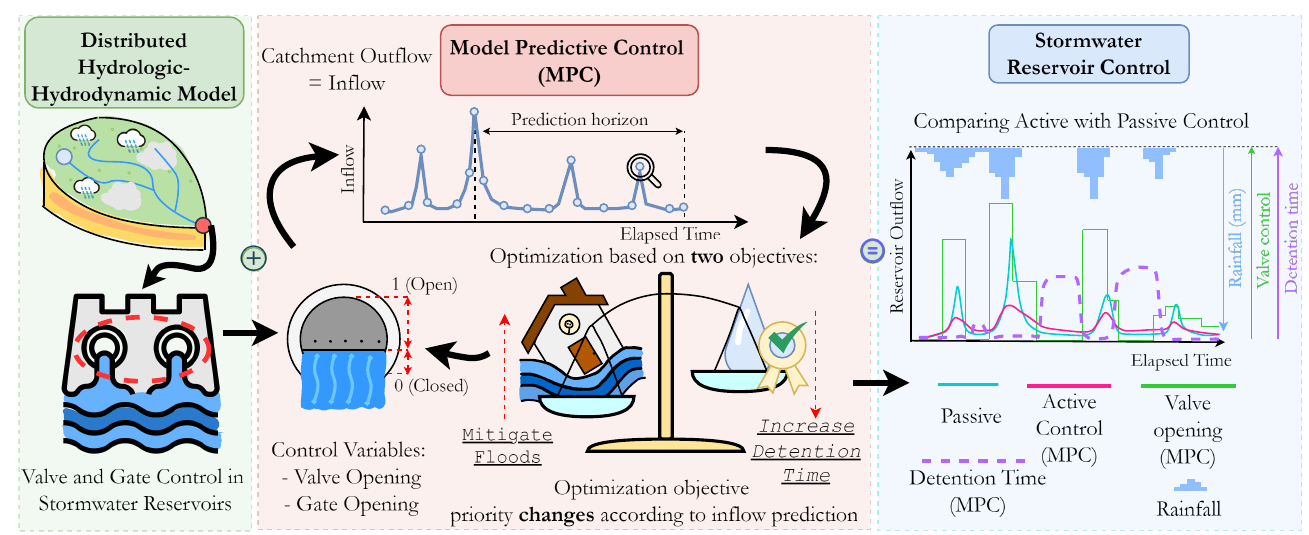}

\section{Introduction}
Floods are becoming more frequent due to urbanization and climate change \citep{miller2017impacts,lu2022spatial,gao2020effects}. Estimates indicate that more than USD 1 trillion between 1980 and 2013 were indirectly associated with flood damages \citep{winsemius2016global}. Severe storms and inland flooding combined caused more than USD 650 billion between 1980 and 2023, only in the United States \citep{smith2024}.  It is expected that new strategies for flood adaptation will be required to cope with more extreme and frequent events in the coming decades. Such strategies include implementing and retrofit stormwater infrastructure such as reservoirs, channels, tunnels, and volume reduction techniques that promote runoff infiltration \citep{zahmatkesh2015low}. 

Retrofitting new infrastructure in urbanized areas is often infeasible or cost-prohibitive, especially in large and dense urban centers \citep{cook2007green} due to the lack of physical space. A common characteristic in large and older cities is the occupation of floodplains and areas along rivers and creeks,  which restrict the implementation of new infrastructure such as online reservoirs or even off-line systems \citep{walsh2001effects}. Alternatively, retrofitting existing stormwater systems with techniques that allow for increased performance without requiring new areas, such as real-time controls, is a viable alternative to mitigate the impacts of climate change and urbanization in stormwater.

Real-time control (RTC) in stormwater reservoirs can provide multiple benefits related not only to flood control \citep{gomes2022flood,wong2018real,sadler2020exploring}, but also to water quality \citep{oh2023model,sharior2019improved} and erosion control \citep{schmitt2020simulation}. RTC works by controlling a physical system (e.g., reservoirs, channels) by changing the flow area of a hydraulic device (e.g., orifice valves, spillway gates) through actuators to establish a controllable operation that typically has only one goal (e.g., peak flow mitigation) \citep{oh2023model}. 

However, proof-of-concept of stormwater RTC applications is yet to be limited in the literature, although an extensive literature exists for Combined Sewer Overflows (CSO) \citep{zhang2023towards,van2023real,van2022towards,castelletti2023model,duchesne2001mathematical,jean2022real}. Often, existing studies of stormwater RTC are limited to modeling studies for flood control only \citep{schmitt2020simulation,wong2018real}, with results for very specific case studies or with limited modeling approaches that might not apply to other real-world cases \citep{webber2022moving}. To address a more generalizable methodology for stormwater RTC in small catchments, a physically-based model that couples RTC techniques in a framework with a quasi 2-D watershed hydrodynamic model and reservoir routing modeling is developed. The RTC technique employs a Model Predictive Control (MPC) algorithm to manage the functioning of valves and gates in stormwater reservoirs to reduce floods and improve water quality.

MPC is a control technique that seeks to optimize a system's performance based on predictions of future states in a feedback loop. In the case of stormwater systems, MPC is usually used associated with weather forecasting that would predict rainfall intensity. By solving multiple optimization problems for each new rainfall forecast, the MPC can define an optimized control strategy that is aligned with the flood control objectives of the system \citep{gomes2022flood}.

By retrofitting existing reservoirs with RTC coupled with MPC techniques, decision-makers, planners, and designers can increase flood performance without requiring new areas to construct reservoirs. With the advent of the Internet of Things (IoT), inexpensive sensors, and free available GIS datasets, one can increase the efficiency of reservoirs by simply applying a valve and gate operating system \citep{wong2018real}.

 In general, the recent literature combines data-driven, artificial intelligence, and ruled-based approaches, with some studies using physics-based modeling and optimization-based algorithms. Recent studies apply RTC using data-driven algorithms such as reinforcement learning and neural networks. The research developed in \citet{mullapudi2020deep} applies a reinforcement learning technique for flood mitigation to control valves in stormwater reservoirs, which requires data observation and a relatively high computational cost and data storage requirement. Similarly, \citet{zhang2018simulation} developed a neural network to estimate turbidity and TSS concentrations, focusing on developing a water quality-based RTC. Both studies do not provide a fully mathematical description of flood and water quality processes and focus on data observation. Although a shift from the computational paradigm towards the data-centric in water engineering is occurring \citep{fu2024making}, the availability of enough data to develop data-driven models can be a problem in developing countries.

Physics-based models coupled control techniques can be a solution for the generalization and wider application RTC for stormwater flood management. The Storm Water Management Model (SWMM), which uses 1D Saint-Venant-Equation solver for channel flood routing, has also been applied in several studies for RTC of stormwater systems. Research conducted in \citet{maiolo2020use} coupled SWMM to simulate movable gates in conduits and reduce peak flows. Other studies, such as those presented in \citet{bartos2021pipedream}, demonstrate the use of digital twins of the catchment for flood modeling and, more recently, for pollutant modeling \citep{kim2024digital}. 

The research presented in \citet{bilodeau2018real} evaluated peak flow reduction and detention time using PCSWMM, and the results obtained show that both objectives can be satisfied with RTC. The study presented in \citep{wong2018real} also coupled the SWMM solver with a reactive control (i.e., a control not based on future predictions) algorithm that controls the valve opening in stormwater reservoirs through reactive optimization control, indicating that the RTC, even when not applied with predictive controllers, can provide better performance than passive cases. The use of MPC as control technique, however, is shown to be more adaptative to a wide range of forecast and can outperform reactive and passive scenarios and most cases \citep{gomes2022flood}.

One issue with implementing RTC in stormwater reservoirs arises when they are connected in a cascade system and the control is centralized per reservoir. Research conducted in \citet{ibrahim2020real} linked this problem and developed a hybrid approach for systems with various reservoirs that can have feedback according to the applied controls. As in previous studies, this research utilized conceptual hydrological models that may be sensitive to different catchments and necessitate thorough data-driven calibration for parameters that are not directly measured.


In addition to the applications of RTC in stormwater systems and its potential benefits, implementing RTC in a real-world system still has some drawbacks. One of the issues of RTC is how municipalities would accept a technology to automatically manage flood control measures, such as valves, gates, and pumps. Research conducted in \citet{naughton2021barriers} showed that municipalities are reluctant to implement RTC for reasons such as operational and maintenance costs. However, the RTC of stormwater reservoirs has been shown to enable stormwater reservoirs to reach 80\% TSS removal in Wisconsin, meeting local water quality criteria \citep{naughton2021barriers}. RTC in stormwater reservoirs coupled with physics-based modeling of the watershed is promising for optimizing reservoir performance for floods and runoff quality treatment \citep{oh2023model}. However, the application of MPC requires a relatively rapid model (i.e., plant model in control theory) of the underlying flood routing and water quality transport and fate systems, which can be very complex due to the non-linear flood routing and infiltration models that might be required to simulate real-world cases. 

It is observed from the aforementioned literature a trend of using simplified plant models (i.e., a typical model that only captures the most significant part of the complete system dynamics of watersheds and reservoirs) to delineate the hydrodynamic and pollutant transport and fate phenomena. Data-driven algorithms and black-box models are also becoming more common with monitoring and data-gathering advances. These techniques are more feasible when high-fidelity and frequent observations of flow discharges, water levels, and pollutant concentrations are available, which is a drawback for poorly monitored stormwater systems. A more accurate description of plant dynamics is developed to be less dependent on field-specific observations and to create a more generalizable method that could be case study-free.

Naturally, some degree of parameter estimation is required, such as soil properties and land roughness; however, these have physical meaning and are relatively simpler to estimate with freely available GIS datasets than site-specific parameters used in a black-box model, for example. The model capabilities are expanded by using the most physics-based parameters possible to allow for generalization and a wider application. However, even though our ability to explain hydrodynamic processes is increasing, simulating water quality without field observations is still complex. 

In the proposed approach, the  \textit{detention time} is used, that is, the time in which stormwater runoff is stored in the reservoir while no inflow is entering, as a proxy metric related to water quality. Relatively higher detention times increase the proliferation of diseases by the action of bacteria, while relatively lower detention times do not allow enough time for the sedimentation of particulate, for example. Thus, one can define optimal \textit{detention time} (i.e., usually between 18-h to 36-h) as an alternative to modeling the complex dynamics of water quality and be able to allow RTC of stormwater runoff pollution by allowing sedimentation while avoiding undesirable biologic treatment. 

With that mentioned, the objectives of this paper are threefold:

\begin{itemize}
    \item Develop and validate an integrated optimization framework modeling of watershed and reservoir dynamics that allows integrated real-time control of water quantity and water quality in flood control reservoirs.
    \item Evaluate how the developed approach would perform against passive scenarios with the valve fully or partially opened for discrete critical events and 1 hydrological year of continuous simulation. 
\end{itemize}

Achieving these objectives leads to the fundamental contributions of this paper, described as follows:

\begin{itemize}
    \item The model developed in \citep{gomes2022flood} is expanded to include a quasi 2-D kinematic-wave state space hydrodynamic model. Valve and gate control are included as control variables and spatial modeling of rainfall and evapotranspiration. Moreover, modeling reservoirs with variable stage-area and stage-porosity is allowed to simulate low-impact development techniques.
    \item A proof-of-concept of the developed model by comparing results with the HEC-RAS 2D full-momentum model to provide a validation scenario for the hydrological simulations of the watershed model is performed.    
    \item An adaptive MPC optimization problem that switches the weights and the objective function according to future predictions of inflows, allowing flood and water quality-based control, is developed.
    \item A novel smart control technique is developed to mitigate both major and minor floods, while also facilitating the removal of pollutants by increasing detention times.
\end{itemize}


\section{Material and Methods} \label{sec:material}

An enhanced version of the real-time control stormwater model (RTC-SM) presented in \citet{gomes2022flood} is developed. Three novel advancements were included and are briefly described here and later detailed in this section. These advancements are important to adapt the previous model \citep{gomes2022flood} to broader case studies where a more detailed water balance modeling is required, in addition to being able to simulate reservoirs with complex bathymetry. First, the watershed model now accounts for evapotranspiration modeling using the Penman-Monteith method \citep{sentelhas2010evaluation}. Second, the model was expanded to include groundwater replenishment using the properties of the uppermost soil layer and saturated hydraulic conductivity to derive replenishing rates \citep{rossman2016storm}. Third, the reservoir model can now account for stage-varying areas, allowing the simulation of real-world natural reservoirs with complex bathymetry.

Using mostly physics-based models, the RTC-SM model solves watershed, reservoir, and 1-D channel routing. The watershed is discretized into finite cells, and flow is assumed to route toward the steepest surface elevation gradient. Infiltration is modeled using the Green-Ampt model \citep{Green1911Studies}, and transformation of water depth into flow is performed using the non-linear reservoir model \citep{Rossman2010}. A gradient boundary condition is assumed at the watershed outlet, which discharges into a stormwater reservoir.

The reservoir can be modeled as either a stage-varying area and volume or as a prismatic reservoir. Moreover, it can be modeled with stage-varying porosity (e.g., porosity varying with depth in case the reservoir has layered soils) or with 100\% void content (that is, regular detention ponds). In other words, the model can be used to simulate real-time control modeling of infiltration Low Impact Developments (LIDs) stormwater control measures such as bioretentions and permeable pavements. Although the model can simulate 1-D channels using the diffusive wave model, this feature is not investigated \citep{gomes2022flood}.

\subsection{Watershed Hydrologic and Hydrodynamic Modeling}
\begin{figure}
    \centering
    \includegraphics[scale = .75]{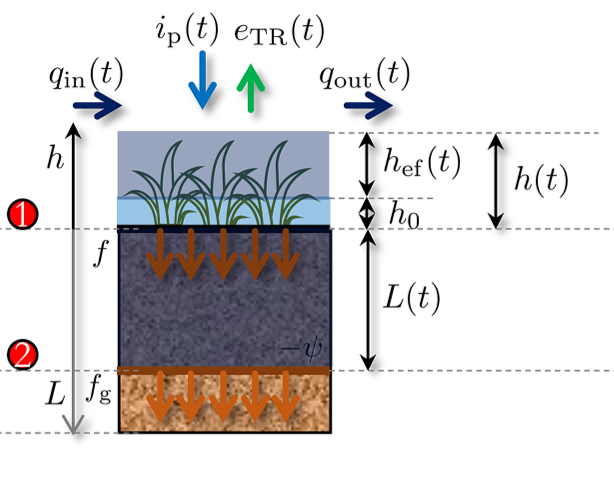}
    \caption{Hydrologic conceptual model where $q_{\mr{in}}$ and $q_{\mr{out}}$ are inflows and outflows [$\mr{L \cdot T^{-1}}$], $i_{\mathrm{p}}$ is the rainfall intensity [$\mr{L \cdot T^{-1}}$], $e_{\mathrm{TR}}$ is the real evapotranspiration [$\mr{L \cdot T^{-1}}$], $f$ is the infiltration rate [$\mr{L \cdot T^{-1}}$], $f_{\mr g}$ is the groundwater replenishing [$\mr{L \cdot T^{-1}}$], $\psi$ is the suction head acting at the wetting front [$\mr{L}$], $\Delta \theta$ is the soil moisture deficit [-], $h_{\mathrm{ef}}$ is the effective water depth [$\mr{L}$] while $h$ is the total depth [$\mr{L}$], $h_{\mr 0}$ are the losses through plant interception and initial abstractions [$\mr{L}$], $L(t)$ is the effective depth of the saturated zone [$\mr{L}$], and $f_{\mathrm{d}}$ is the cumulative infiltrated depth [$\mr{L}$]. Two fundamental equations are solved for the atmosphere-soil (1) interface and the wetting front-soil interface (2). }
    \label{fig:conceptual}
\end{figure}

The overall mass balance differential equation is written for all cells of the domain and accounts for the surface and the groundwater replenishing mass balance as follows (see Fig.~\ref{fig:conceptual}):

\begin{subequations}
\begin{align}
    \frac{\partial \m h_{\mr{ef}}(t)}{\partial t}&={\m q_{\mr{in}}}(t)- {\m q_{\mr {out}}}(t) + \m i_\mr{p}(t) - \m e_{\mr{TR}}(t) - \m f(t) \\
    \frac{\partial \m f_{\mr d} (t)}{\partial t} &= \m f(t) - \m f_{\mr{g}},
\end{align}
\end{subequations}

The water balance in the watershed cells is calculated from inflows, outflows, rainfall, evapotranspiration, and infiltration. For a given cell ($i,j$) in the catchment domain, using a finite-difference forward Eulerian scheme, the mass balance is discretized into:

\begin{subequations} 
\begin{align} \label{equ:finite_diff}
     h^{\mr{i,j}}_{\mathrm{ef}}(k+1) &=  h^{\mr{i,j}}_{\mathrm{ef}}(k) + \Delta t\Bigl(q^{\mr{i,j}}_{\mathrm{in}}(k) + i^{\mr{i,j}}_\mathrm{p}(k) - e^{\mr{i,j}}_{\mathrm{TR}}(k)  -  q^{\mr{i,j}}_{\mathrm{out}}(k) - f^{\mr{i,j}}(k) \Bigr) \\
    f^{\mr{i,j}}_{\mr d}(k+1) &= f^{\mr{i,j}}_{\mr d}(k) + \Delta t \Bigl(f^{\mr{i,j}}(k) - f^{\mr{i,j}}_{\mr g} \Bigr),
\end{align}
\end{subequations}
\noindent where $\Delta t$ = model time-step [$\mr{T}$], $k$ = time-step index [-].

The full derivation of the hydrological inputs, such as evapotranspiration, Green-Ampt \citep{Green1911Studies} infiltration modeling, and groundwater replenishment rate, is described in Supplemental Information (SI).

To solve Eq.~\eqref{equ:finite_diff}, $\m q_{\mr{in}}$ and $\m q_{\mr{out}}$ are calculated from a momentum equation. Manning's equation is used as flow-depth relationship coupled with the non-linear reservoir method to allow modeling of losses to initial abstraction, such that:

\begin{subequations} \label{equ:manning}
\begin{align}
q_{\mr x}^{\mr{i, j}}=\frac{\max\Bigr({h^{\mr{i,j}} - h_0^{\mr {i,j}},0}\Bigr)^{5 / 3}}{n^{\mr{i, j}}}\left( s^{\mr {i,j}}_{\mr f}\right)^{1 / 2} \Delta y, \\
q_{\mr y}^{\mr{i, j}}=\frac{\max\Bigr({h^{\mr{i,j}} - h_0^{\mr {i,j}},0}\Bigr)^{5 / 3}}{n^{\mr{i, j}}}\left( s^{\mr {i,j}}_{\mr f}\right)^{1 / 2} \Delta x,
\end{align}
\end{subequations}
where  $s_{\mr f}$ is the friction slope, assumed as the bottom slope, $n$ is the Manning’s roughness coefficient [$\mr{T \cdot L^{-1/3}}$], $\Delta x$ is the pixel $x$ discretization [$\mr{L}$], and $\Delta y$ is the pixel $y$ discretization [$\mr{L}$]. From now on a raster flood model with $\Delta x = \Delta y$ is assumed. 

The kinematic wave approach is computationally faster than diffusive \citep{gomes2023hydropol2d} local-inertial \citep{bates2010simple}, and full dynamic models \citep{Brunner}; however, it is not appropriate in complex terrain with low relief or where phenomena such as backwater effects govern the hydrodynamics \citep{neal2012subgrid}. Besides these limitations, for catchments with relatively steep terrain gradients, the kinematic-wave approximation presents accurate results \citep{yu2014high}. 

From the previous equation, it is observed that the hydraulic radius is assumed to be the effective depth of the water ($h - h_0$), which is more applicable in shallow water approximations when smaller depths and larger grid dimensions are used in the simulation \citep{akan2021open}. The friction slope used in Eq.~\eqref{equ:manning} must be positive and directed toward the central cell's downstream cell, assuming all cells have an assigned flow direction. 

The mass balance equation shown in Eq.~\eqref{equ:finite_diff} requires the calculation of intercell flows and a flow direction matrix for the $x-x$ and $y-y$ directions. The procedure to derive these matrices is detailed in the SI. In addition, the procedure to estimate the adaptive time-step based on the Courant-Friedrichs-Levy (CFL) \citep{courant1928partiellen} is presented in the SI.




\subsection{Reservoir Model}
The reservoir receives inflow from the watershed, which can be calculated by solving  Eq. \eqref{equ:manning} for the outlet cell $\m i_0$. Therefore, the reservoir net inflow can be calculated as:

\begin{equation} \label{equ:watetrshed_inflow}
    q_{\mr{in}}^{\mr r}(k) = q_{\mr{out}}^w(k) + \Bigl(i_{\mr p}^{\mr r}(k) - e_{\mr p}^{\mr r}(k) \Bigr) \omega_{\mr r} (k),
\end{equation}
where $q_{\mr{out}}^w$ is the watershed outflow [$\mr{L \cdot T^{-1}}$], $i_{\mr p}^{\mr r}$ [$\mr{L \cdot T^{-1}}$] is the rainfall intensity in the reservoir, $e_{\mr p}^{\mr r}$ [$\mr{L \cdot T^{-1}}$] is the evaporation, and $\omega_{\mr r} := f(h^{\mr r}(k))$ is the depth-varying reservoir area [$\mr{L^2}$]. 

The derivation of $\omega_{\mr r}$ is made from known stage x area values. A continuous function of the reservoir area in terms of water depth is built for each known stage x area value. A detailed explanation and mathematical derivation of this procedure is shown in the SI. For simplicity of notation, assume $\omega_{\mr r}(h) = \omega_{\mr r}$ and $\eta(h) = \eta$.

By solving a mass balance equation in Eq.~\eqref{equ:mass_balance_reservoir} and an energy equation in Eq.~ \eqref{equ:outflow_reservoir}, one can determine the reservoir outflow as follows \citep{gomes2022flood}:

\begin{equation} \label{equ:mass_balance_reservoir}
    h^{\mr r}(k+1)=h^{\mr r}(k)+\Bigl(\frac{\Delta t}{\omega^{\mr r} \eta} \Bigl)\left(q_{\mathrm{\mr{in}}}^{\mr r}(k)-q_{\mathrm{out}}^{\mr r}(k)\right)
\end{equation}

\begin{equation} \label{equ:outflow_reservoir}
    q_{\text {out }}^{\mr r}\Bigl(h^{\mr r}(k), u_{\mr v}(k), u_{\mr s}(k)\Bigr)=\left\{\begin{array}{l}
u_{\mr v}(k) k_{\mr o} \Bigl(\widehat{h}^{\mr r}_{\mr v}(k)\Bigr)^{\alpha_{\mr v}} \text { if } h^{\mr r}(k) \leq p, \quad \text { else } \\
u_{\mr v}(k) k_{\mr o} \Bigl(\widehat{h}^{\mr r}_{\mr v}(k)\Bigr)^{\alpha_{\mr v}}+ u_{\mr s}(k) k_{\mr s}\Bigl(\widehat{h}^{\mr r}_{\mr s}(k)\Bigr)^{\alpha_{\mr s}},
\end{array}\right.
\end{equation}
where $k_{\mr o}$ [$\mr{L^{3 - \alpha_v}\cdot T^{-1}}$] and $k_{\mr s}$ [$\mr{L^{3 - \alpha_s}\cdot T^{-1}}$] are the orifice and spillway linear rating-curve coefficients, and $\alpha_{\mr o}$ [-] and $\alpha_{\mr s}$ [-] are the exponents of these rating-curves. The variable $\widehat{h}^{\mr r}_{\mr v}\left(h^{\mr r}(k)\right)=\max \left(h^{\mr r}(k)-\left(h_0+h_{\mathrm{m}}\right), 0\right)$, is the effective water depth at the orifice [$\mr{L}$], $h_0$ is orifice depth from the bottom [$\mr{L}$], $h_m = 0.2d_h$, $d_h$ is the hydraulic diameter of the orifice [$\mr{L}$], $\widehat{h}^{\mr r}_{\mr s}\left(h^{\mr r}(k)\right)= \max(h^{\mr r}(k) - p,0)$, is the effective water depth at the gate [$\mr{L}$], where $p$ is the spillway elevation [$\mr{L}$],  $\omega^{\mr r} := f\left(h^{\mr r}(k)\right)$ is the reservoir stage-area function [$\mr{L^2}$], $u_{\mr v}$ is the valve openness [-] and $u_{\mr s}$ [-] is the gate openness. 

By calculating the Jacobians of Eq.~\eqref{equ:outflow_reservoir}, one can derive a linearized model for the reservoir outflow that considers control in valves and gates by performing a Taylor's series 1st-order approximation. Let $\alpha(k)$ be the Jacobian of Eq.~\eqref{equ:outflow_reservoir} with respect to $h^{\mr r}$, $\beta(k)$ the Jacobian with respect to $u_{\mr v}$, and $\gamma(k)$ the Jacobian with respect to $u_{\mr s}$, presented as follows:

\begin{subequations} \label{equ:linearizations}
\begin{align}
\overbrace{\frac{\partial q_{\mathrm{out}}^{\mr r}\left(h^{\mr r}(k), u_{\mr v}(k), u_{\mr s}(k)\right)}{\partial h^{\mr r}}}^{\alpha(k)} &= \alpha_{\mr v} u_{\mr v} k_{\mr o} \Bigl(\widehat{h}_{\mr v}(k) \Bigr)^{\alpha_{\mr v} - 1} + \alpha_{\mr s } u_{\mr s}(k) k_{\mr s} \Bigl(\widehat{h}_{\mr s}(k) \Bigr)^{\alpha_{\mr s} - 1} \\
\overbrace{\frac{\partial q_{\mathrm{out}}^{\mr r}\left(h^{\mr r}(k), u_{\mr v}(k)\right)}{\partial u_{\mr v}}}^{\beta(k)} &= k_{\mr o} \Bigl(\widehat{h}_{\mr v}(k) \Bigr)^{\alpha_{\mr v}} \\
\overbrace{\frac{\partial q_{\mathrm{out}}^{\mr r}\left(h^{\mr r}(k), u_{\mr s}(k)\right)}{\partial u_{\mr s}}}^{\gamma (k)} &= k_{\mr o} \Bigl(\widehat{h}_{\mr s}(k) \Bigr)^{\alpha_{\mr s}},
\end{align}   
\end{subequations}
which ultimately turns out in a linearized model around an equilibrium point $\m x_{\mathrm{eq}} = [h^{\mr r}_*, u_{\mr v}^*, u_{\mr s}^*]^{\mr{T}}$ that collects the reservoir depth, valve and gate openness as follows:

\begin{equation} \label{equ:linearized_outflow}
\begin{aligned}
q_{\mathrm{out}}^{\mr r}\left(h^{\mr r}(k), \m x_{eq}\right) &= \overbrace{q_{\mathrm{out}}^{\mr r}\left(h_*^{\mr r}\right)}^{\epsilon(k)}+\overbrace{\left.\alpha\right|_{h^{\mr r}=h_*^{\mr r}, u_{\mr v} = u_{\mr v}^*, u_{\mr s} = u_{\mr s}^*}}^{\tilde{\alpha}(k)}\left(h^{\mr r}(k)-h_*^{\mr r}\right)   \\
&+ \overbrace{\left.\beta\right|_{h^{\mr r}=h_*^{\mr r}, u_{\mr v} = u_{\mr v}^*}}^{\tilde{\beta}(k)}\left(u_{\mr v}(k)-u_{\mr v}^*\right) + \overbrace{\left.\beta\right|_{h^{\mr r}=h_*^{\mr r}, u_{\mr v} = u_{\mr v}^*}}^{\tilde{\gamma}(k)}\left(u_{\mr s}(k)-u_{\mr s}^*\right),
\end{aligned}
\end{equation}
where the orifice is defined by parameters $k_{\mr o}$ and $\alpha_{\mr v}$ and the gate is described by $k_{\mr s}$ and $\alpha_{\mr s}$. 

Therefore, tunning the orifice and gate parameters with different values allows the model to simulate rectangular, circular, or variable shape orifices while gates can be modeled as open spillways (i.e., Thompson Spillway $\alpha_{\mr s} = 3/2 $) or as controllable gates simulated as orifices (i.e., $\alpha_{\mr s} = 1/2$). The linear parameters $k_{\mr o}$ and $k_{\mr s}$ are the multiplication of all linear terms and terms inside exponential equations that are not a function of the representative water depth in the hydraulic equations. For the orifice case, $k_{\mr o} = c_d a_{ef} \sqrt{2g}$, where $c_d$ is the orifice discharge coefficient [-], $a_{ef}$ is the orifice area [$\mr{L^2}$], and $g$ is the gravity acceleration [$\mr{L\cdot T^{-2}}$]. More details on the modeling of the governing coefficients can be found in \citep{gomes2022flood,french1985open}.

Substituting the inflow discharge from the watershed [Eq.~\eqref{equ:watetrshed_inflow}], which is input data for the reservoir model derived from the watershed model, the linearized outflow discharge [Eq.~\eqref{equ:linearized_outflow}] derived from Eqs.~\eqref{equ:outflow_reservoir} and \eqref{equ:linearizations}, and tracking the outflow discharge as the output of the reservoir model, a reservoir state-space model is developed, such that:

 \begin{equation} \label{equ:state_space}
  \begin{bmatrix}
      h^{\mr r}(k+1) \\ q_{\mr{out}}^{\mr r}(k) 
  \end{bmatrix}
  =
  \begin{bmatrix}
      A(k) \\
      C(k)
  \end{bmatrix}
    h^{\mr r}(k)
    +
    \begin{bmatrix}
        B^{\mr v}(k) \\
        D^{\mr v}(k)
    \end{bmatrix}
    u_{\mr v}(k)
    +
    \begin{bmatrix}
        B^{\mr s}(k) \\
        D^{\mr s}(k)
    \end{bmatrix}
    u_{\mr s}(k)
    +
    \begin{bmatrix}
        \phi(\m x_{\mr{eq}}(k)) \\
        \epsilon(\m x_{\mr{eq}}(k))
    \end{bmatrix}
 \end{equation}
where the tracked state is the reservoir water depth $(h^{\mr r}(k))$ and the output is the reservoir outlet discharge $(q_{\mr{out}}^{\mr r})$ being controlled by the valve ($u_{\mr v}$) and gate ($u_{\mr s}$) openings. The first row represents a mass balance equation, whereas the second row is an energy balance equation. Time-varying single matrices $A(k)$, $B^\mr{v}(k)$, $B^\mr{s}(k)$, $C^{\mr v}(k)$, $C^{\mr s}(k)$, $D^{\mr v}(k)$, $D^{\mr s}(k)$ are derived taken the linear terms of $h^{\mr r}$ or $u_{\mr v}$, or $u_{\mr s}$ from Eqs.~ \eqref{equ:mass_balance_reservoir} and \eqref{equ:outflow_reservoir}. The source and linearized terms $\phi(k)$ collect the inflow discharge from the watershed, and the operational point substitution results in the linearized outflow discharge equation in Eq.~\eqref{equ:outflow_reservoir}. Similarly, $\epsilon(k)$ collects the application of the operational point into the linearized terms of the discharge, in Eq.~\eqref{equ:outflow_reservoir}.

The previous derivation of the state-space model in Eq.~\eqref{equ:state_space} can be easily expanded to more reservoirs and watersheds using topological relationships \citep{gomes2022flood}.

\subsection{Model Predictive Control}

\subsubsection{Objective Function}
The operation of stormwater reservoirs can have multiple goals. For flood control, reservoirs should be operated to mitigate peak flows. One way to attenuate peak flow using active control techniques is to transform the reservoir operation into a minimization of flood costs \citep{gomes2022flood}. Flood costs can be either the cost of flood damage or can be abstracted to include various flood-related costs such as (i) valve operation (i.e., control energy), (ii) maximum water level at the reservoir, or (iii) violation of maximum tolerable outflow. These objectives can be grouped into a single-objective function given by:
 
\begin{align} \label{equ:optimization_problem} 
\min _{\boldsymbol{u}_{\mathrm{v,k}}, \boldsymbol{u}_{\mathrm{s,k}}} \sum_{k=0}^{N_{\mr p}-1} & \boldsymbol{J}\Bigl(\boldsymbol{h}^{\mr r}(k+1), \boldsymbol{u_{\mr v}}(k+1), \boldsymbol{u_{\mr s}}(k+1)\Bigr) =  \rho_{\mathrm{u}}\left\|\Delta \boldsymbol{u}_{\mathrm{v,k}}^{\mr{T}}\right\|_2^2  
+ \rho_{\mathrm{u}}\left\|\Delta \boldsymbol{u}_{\mathrm{s,k}}^{\mr{T}}\right\|_2^2   \notag \\ & + \rho_{r}\Bigl( \left\| \max{(\m h^{\mr r} - h_{\mr{ref}}^{r},0)}\right\|_{\infty}\Bigr) +\rho_{\mathrm{q}, \mathrm{k}}\left(\left\|\max \left(\boldsymbol{q}_{\mathrm{k}}^{\mathrm{r}} -{q}^{\mr r}_{\mathrm{ref}, \mathrm{k}} ,0\right)\right\|_{\infty}\right) \notag \\
&+ \rho_{**}\Bigl(\left\| \max{(\m q_k^{\mr r} - q_{max}^{**},0)}\right\|_{\infty}\Bigr),
\end{align}
where $\m J$ is the cost function [-] and their weights are given for the control input $(\rho_{\mr u})$, for the depths of the surface of the water in the reservoirs $(\rho_{\mr r})$ and for the exceedance of the maximum tolerable flow $(\rho_{\mr q})$. $N_{\mr p}$ is the prediction horizon, $\Delta \m u_{\mr{v,k}} = [\Delta u_{\mr{v,1}} \dots \Delta u_{\mr{v,N_{{p-1}}}}]^{\mr{T}}$, $\Delta \m u_{\mr{s,k}} = [\Delta u_{s,1} \dots \Delta u_{\mr{s,N}_{\mr{p-1}}}]^{\mr{T}}$$, \m h_{\mr{k}}^{\mr r} = [h^{\mr r}_{\mr 1}, \dots h^{\mr r}_{\mr{N_{\mr p} - 1}}]^{\mr{T}}$, $\m q^{\mr r} = [ q_{\mr{out,1}}^{\mr r}, \dots  q_{\mr{out,N_{\mr p} - 1}}^{\mr r}]^{\mr{T}}$, $\m q_{\mr{ref}}^{\mr r}$(k) is the time-varying reference outflow. 

\begin{figure}
    \centering
    \includegraphics[scale = 0.23]{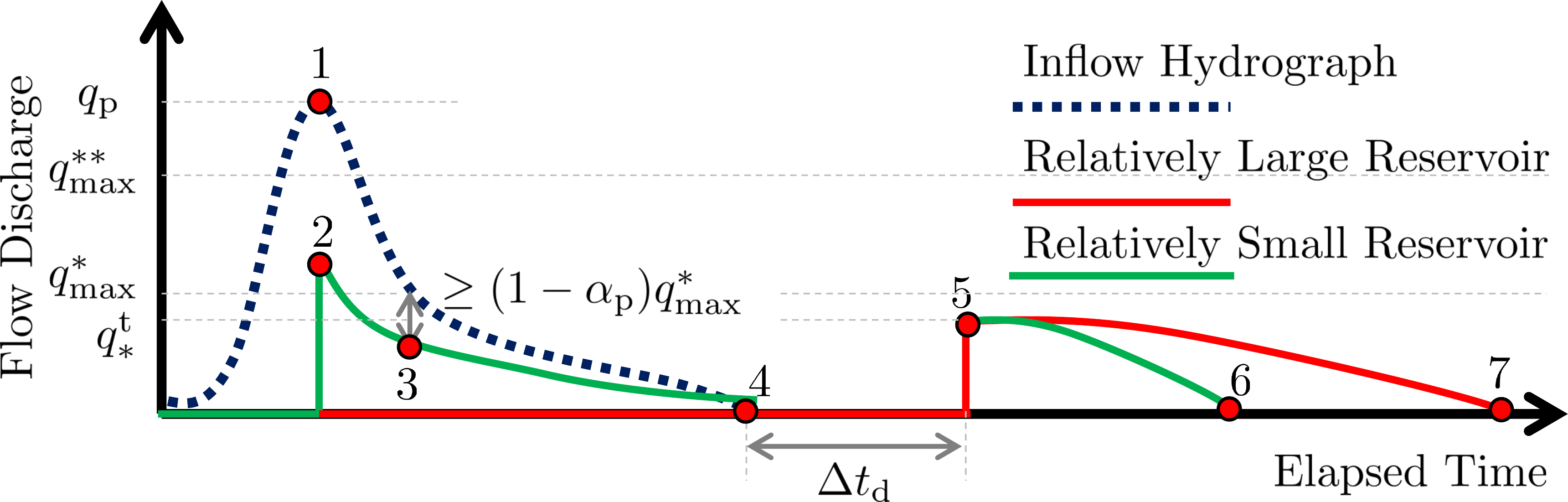}
    \caption{Inflow and outflow hydrographs of a relatively large and relatively small stormwater reservoir with water quantity and quality control, where $\Delta t_{\mr d}$ is the required detention time. The factor $\alpha_{\mr p}$ can be tuned and be used to represent a desired peak flow reduction under minor flood events with maximum predicted inflows smaller or equal than $q_{\mr{max}}^{*}$. The relatively large reservoir can store all inflow hydrograph and later release after a detention time threshold is reached, while the relatively small reservoir does not have the storage capacity to do so and has to be operated focusing on flood mitigation. }
    \label{fig:schematics}
\end{figure}

The problem constraints are given by the physics of the system dynamics, and the control signals $\m u_{\mr k}$ have each entry between 0 and 1, such that the valves can only be fully or partially opened. Therefore, the solution of the minimization function of Eq.~\eqref{equ:optimization_problem} is mathematically constrained by:

\begin{subequations} ~ \label{equ:constraints}
\begin{align}
\textrm{subject to:}~~& \mr{Eq.}~\eqref{equ:state_space}\\
  & \Delta{}u_{\mr{min}}\leq{}\Delta u\left(k\right)\leq  \Delta{}u_{\mr{max}} \\
& \m u_{\mr v}(k), \m u_{\mr s}(k)\in{} \mathcal{\m U} := \mathbb{R} \in [0,1] \\
& \m h^{\mr r}(k)\in{} \mathcal{\m X} := \mathbb{R} \in [0,h^{\mr r}_{\mr{max}}]
\end{align}
\end{subequations}
where $h^{\mr r}_{\mr {max}}$ is the maximum reservoir depth [$\mr{L}$].

\begin{figure} 
    \centering 
    \includegraphics[scale = 0.25]{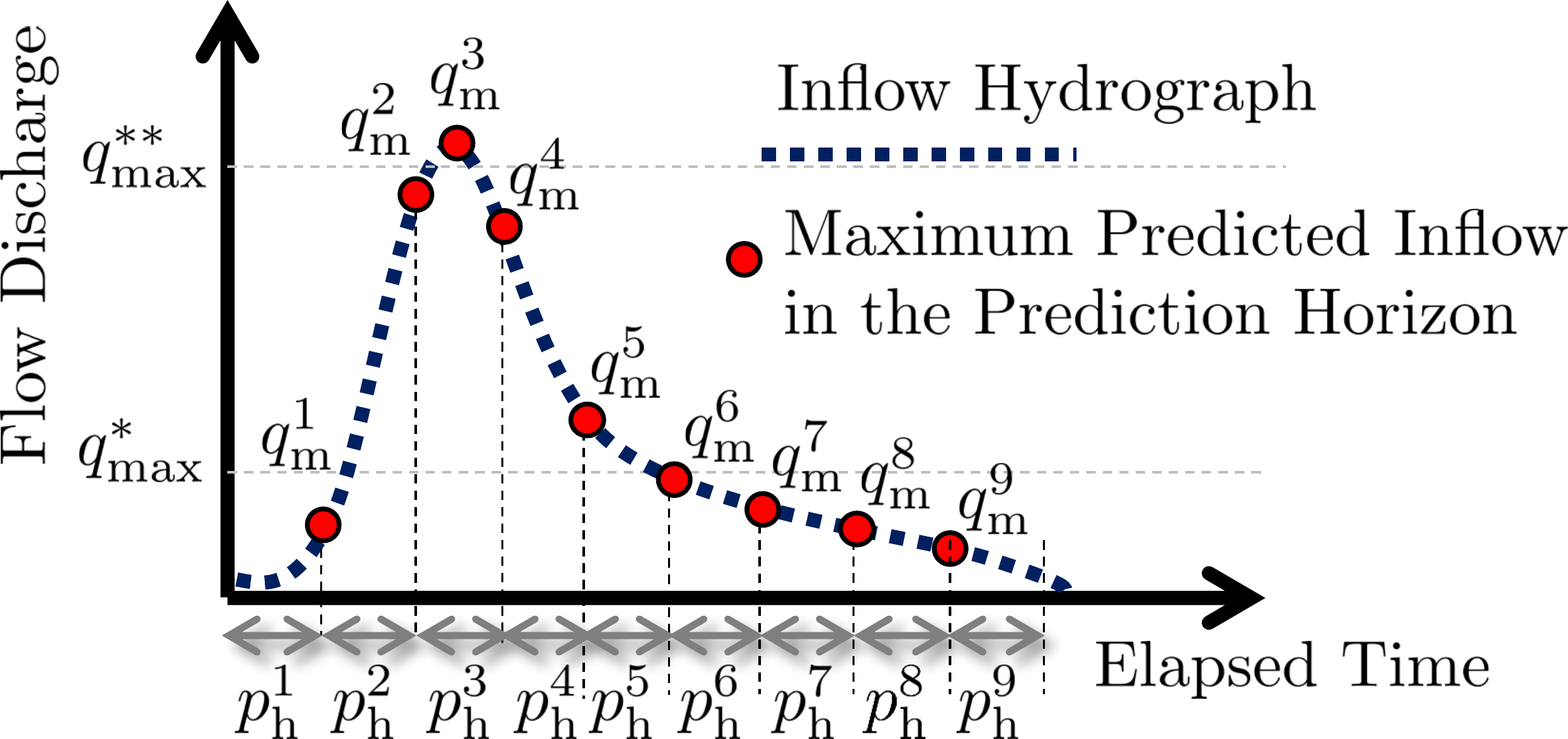}
    \caption{Inflow hydrograph and the definition of the maximum inflow $q_{\mr m}$ for each prediction horizon $p_{\mr h}$. The weights to be used in Eq.~\eqref{equ:optimization_problem} are defined by each maximum inflow $q_m$ for each prediction horizon as shown in Eq.~\eqref{equ:weights}. Therefore, the control theoretical goal is changed according to the predicted flood magnitude. Note that $q_{\mathrm{m}}^3$ violates the threshold for large flood $q_{\mathrm{max}}^{**}$, increasing the focus of the control for flood mitigation, while $q_{\mathrm{max}}^1$, for example, has a maximum predicted inflow in the prediction horizon smaller than the threshold for minor floods. }
    \label{fig:hydrograph_prediction}
\end{figure}

The reference outflow $q_{\mr{ref}}^{\mr r}(k)$ [$\mr{L^3 \cdot T^{-1}}$] represents the operational tolerable flow. A heuristic approach to change the reference outflow according to the maximum inflow ($\m q_{\mr{in,k}}^{\mr r}$) predicted in the control horizon $k$ is designed. The goal is to define a control that could work for large storms and minor events. The reference $\m q_{\mr{ref}}^{\mr r}$ changes according to the predicted inflows and can be written as:

\begin{subequations} \label{equ:weights}
\begin{align} 
    \m q_{\mr{ref,k}}^{\mr r} &= 
    \begin{cases}
        \alpha_{\mr p} \max{(\m q_{\mr{in,k}}^{\mr r})},~\text{if $\max{(\m q_{\mr{in,k}}^{\mr r})} \leq  q_{\mr {max}}^{*}$} \\
        q_{\mr{max}}^{*},~\text{Elsewhere} \\
    \end{cases} \\
         \rho_{\mr{q,k}} &= 
    \begin{cases}
        10^1 \rho_{\mr u},~\text{if $\max{(\m q_{\mr{in,k}}^{\mr r})} \leq  q_{\mr{max}}^{*}$} \\
       10^2\rho_{\mr u},~\text{Elsewhere} \\
    \end{cases} \\
    \rho_{**} &= 
    \begin{cases}
         10^3 \rho_{\mr u},~\text{if } \max{(\m q_{\mr{in,k}}^{\mr r}) \geq q_{\mr {max}}^{**}} \\
         0,~\text{Elsewhere}
    \end{cases}
\end{align}    
\end{subequations}
\noindent where $\alpha_{\mr p}$ represents the tolerable maximum reservoir outflow peak /  maximum inflow from the upstream catchment at the control horizon. In other words, ($1 - \alpha_{\mr p}$) represents the minimum peak flow reduction goal applied only in cases where the maximum predicted inflow on the control horizon is smaller than $q_{\mr{max}}^{**}$. If the predicted maximum inflow is greater than $q_{\mr{max}}^{**}$, one can assume that this is a large event and limit the maximum outflow to $q_{\mr{max}}^{**}$ instead of trying to reduce only a percentage of the maximum flow. An illustrative example of the maximum flows predicted in a prediction horizon is shown in Fig.~\ref{fig:hydrograph_prediction}.

The aforementioned parameters are defined according to the reservoir goals and can be parametrized for different reservoirs, watersheds, and local regulations of maximum outflows. The problem is formulated as a non-linear, non-convex optimization problem by choosing these constraints and details. In this model, two solvers for the solution of Eq.~\eqref{equ:optimization_problem} were implemented, the $\mathtt{patternsearch}$ and the $\mathtt{fmincon}$ solvers. The $\mathtt{fmincon}$ algorithm from Matlab. Previous modeling results using $\mathtt{globalsearch}$ and $\mathtt{genetic~algorithms}$ resulted in overly expensive time solutions and were not utilized in this investigation.


Instead of starting the optimization with randomly multi-start points focusing on finding global minima, initial points are created based on simple ruled-based logic. The rationale behind this is to start with initial points with no valve operation (i.e., no control effort), but that would explore the whole decision space. Given several random inputs ($n_{\mr r}$) for the multi-start search, a series of inputs $\m U_{\mr{0,k}} = [\m u_0^1, \dots \m u_0^{n_{\mr r}}]^{\mr{T}}$ are created, with $\m u_0^i = \frac{i}{n_{\mr r}}[\m 1_{N_{\mr p}\times 1}]~\forall~i \in [1,n_{\mr r}] \in \mbb{N_{++}}$, and the optimization problem is run for each of them. Afterward, the model chooses only the solution with a smaller objective function value, given by Eq.~\eqref{equ:optimization_problem}.

Finally, to accomplish proxy water quality goals (i.e., increase detention time), a routine that identifies the maximum inflow in the prediction horizon is developed, and if this quantity equals 0, the detention time starts to be counted. At the beginning of this phase, the valves are closed (that is, $u_{\mr{v}}(k) = 0$), and if no other event occurs during a maximum detention time period ($\Delta t_{\mr d}$), the outflow is released by opening the valves at a capacity equal to $q_{\mr{t}}^*$. This can be the threshold for minor flood events or can also be tuned as a flow used to avoid erosion, for example. By limiting the outflow to $q_{\mr{t}}^*$, the valve opening can be calculated as:

\begin{equation} \label{equ:control_detention}
    u_{\mr r} = \min{\Bigr(\frac{q_{\mr{t}}^*}{k_{\mr o} \sqrt{h_{\mr r}^{e}}},1\Bigr)},
\end{equation}
where $k_{\mr o} = c_{\mr d} a_{\mr{ef}} \sqrt{2g}$ is the orifice coefficient, $c_{\mr d}$ is the discharge coefficient, $a_{\mr{ef}}$ is the effective area of the orifice, $g$ is the acceleration of gravity and $h_{\mr r}^{\mr e}$ is the reservoir water depth at the end of the detention time.

During wet weather periods, that is, periods where the inflow is smaller than a pre-defined flow threshold (i.e., typically $< 2~\mr{m^3\cdot s^{-1}}$ depending on the catchment area) during the prediction horizon, the MPC stops algorithm and switch the problem to a fully water quality-based control. However, if some of the predicted inflows are positive, the algorithm returns to flood-based control by seeking the minimization of Eq.~\eqref{equ:optimization_problem}. This simple heuristic rule allows us to change the problem control from a flood-based control approach during wet weather events to a water quality-based control approach during dry weather periods while avoiding releasing high flows by limiting the valve opening up to a condition where a maximum flow released that is smaller than $q_{\mr{max}}^*$ is expected.

A conceptual example of two reservoirs receiving the same inflow hydrograph is shown in Fig.~\ref{fig:schematics} to illustrate the idea of the MPC approach with proxy water quality control. In Fig.~\ref{fig:schematics}, each notable point is described as follows. Point 1 represents the maximum inflow peak discharge, which is larger than the threshold for large flood events $q_{\mr{max}}^{**}$. The smaller reservoir cannot hold the total inflow hydrograph volume and has to start to release flows in 2 to allow a desired peak flow mitigation in 3; that is, the maximum peak flow released in 3 follows the desired peak flow factor for minor floods $\alpha_{\mr p}$. This factor is tuned to allow peak flow mitigation under relatively small events, which is a drawback of passive reservoirs designed only for relatively large events and, hence, with relatively large orifices that cannot mitigate small inflow discharges. After the inflow hydrograph stops in 4, both reservoirs have closed the orifice valves. The larger reservoir, however, always had the valves closed since it had enough capacity to store the hydrograph volume. 

After reaching the detention time $\Delta t_{\mr d}$ with no predicted inflow hydrographs in this period, both reservoirs now open the valves, releasing a maximum flow $q_{*}^{\mr t}$, also tuned to represent a desired outflow rate that can be designed to avoid erosion or regulate a minimum flow discharge. The smaller reservoir has a faster stage-area function, providing a larger variation in the depth with a relatively smaller variation in volume, which explains the faster release of the flow, while the larger reservoir takes longer to release all flow hydrograph. Both cases show how the designed MPC approach can enhance flood dynamics.

\subsubsection{Performance Indicators}
To evaluate the performance of the MPC strategy, active-controlled results are compared with a baseline scenario corresponding to the passive scenario, that is, the spillway gate and the valve are fully open. To quantify peak flow mitigation, duration curves corresponding to the frequency of discharges and depths are derived, and the average flow discharges and the root mean squared water depths are calculated during a 1-yr continuous simulation.

For water quality, the detention time and the released volume through the valve are tracked to calculate the average detention time and a proxy representation of the treated runoff volume as a function of the product between the runoff volume of the outflow and the detention time. In summary, the average detention time can be defined as the weighted average between the runoff volume released by the valve and the detention time associated with the release. The treated (i.e., the volume that passed through the bottom orifice) is given by:


\begin{equation} \label{equ:volume_treated}
    V_{\mr {out}}(k) = \int_{0}^kq_{\mr{out}}^{\mr r}(t)\mr{dt} = \sum_{i = 0}^{k} q_{\mr{out}}^{\mr r}(i) \Delta t
\end{equation}

By calculating the product between the runoff volume and the detention time and dividing by the runoff volume, one can derive the average detention time calculated as:

\begin{equation} \label{equ:average_detention_time}
    \bar{\Delta t_{\mr {d}}}(k) = \frac{\sum_{i = 0}^{k} q_{\mr{out}}^{\mr r}(i) \Delta t_{d}(i)}{V_{\mr {out}}(k)}
\end{equation}
where $\bar{\Delta t_{\mr d}}(k)$ is the average detention time for step $k$.

In a theoretically large reservoir where flooding is not a concern, the average detention time will be equal to the expected detention time $\Delta t_d$. The controller is designed to prevent floods, so it may switch from quality-based control to flood control if inflows are predicted before $\Delta t_d$, which can reduce the average detention time.

The detention time is only tracked and updated when the prediction inflow hydrograph within the control horizon is null, and the reservoir water depth is greater than $0.2 d_{\mr{h}}$. In other words, the detention time is only calculated when no flows are predicted, and the water depth is enough to begin to be released by the reservoir passively. In cases of predicting inflow hydrographs or with a water depth stored in the reservoir, the detention time is constantly updated. Following this idea, the treated volume $V_{\mr{out}}$ from Eq.~\eqref{equ:volume_treated} is also only calculated when these conditions are met.

\subsection{Study Area}

\begin{figure}
    \centering
    \includegraphics[scale = 0.45]{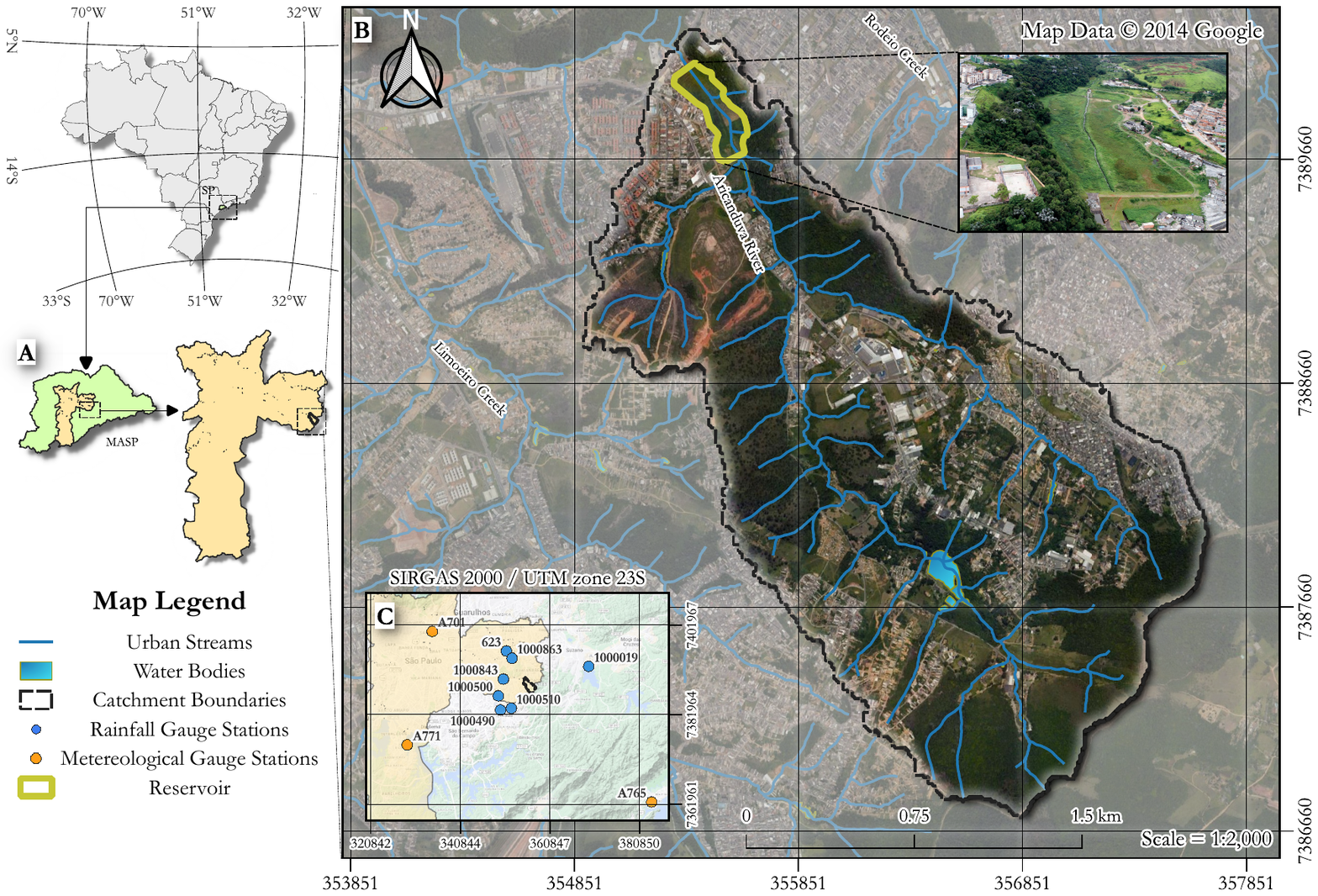}
    \caption{Study Area Map in São Paulo City - Brazil. The reservoir has an inlet channel and receives headwater from the Aricanduva watershed. During large events, runoff is spilled by a rectangular crest spillway.}
    \label{fig:BHARI}
\end{figure}

The Aricanduva-I watershed drains 4.7 km² from the headwaters of the Aricanduva River. Since the urbanization of the São Paulo city, this area has suffered many problems due to extreme hydrological events. A detention reservoir (on-line) responsible for storing $\mathrm{200,000~m^{3}}$ of stormwater runoff (normal capacity below the spillway) discharges through a rectangular orifice (1.0~m $\times$ 1.0~m) operated passively, that is, no orifice control is currently implemented. An emergency spillway can be used during large events, discharging the overflow directly into the downstream drainage system \citep{CadernoBHARI}, as presented in Fig. \ref{fig:BHARI}. 

\begin{figure}
    \centering
    \includegraphics[scale = 0.47]{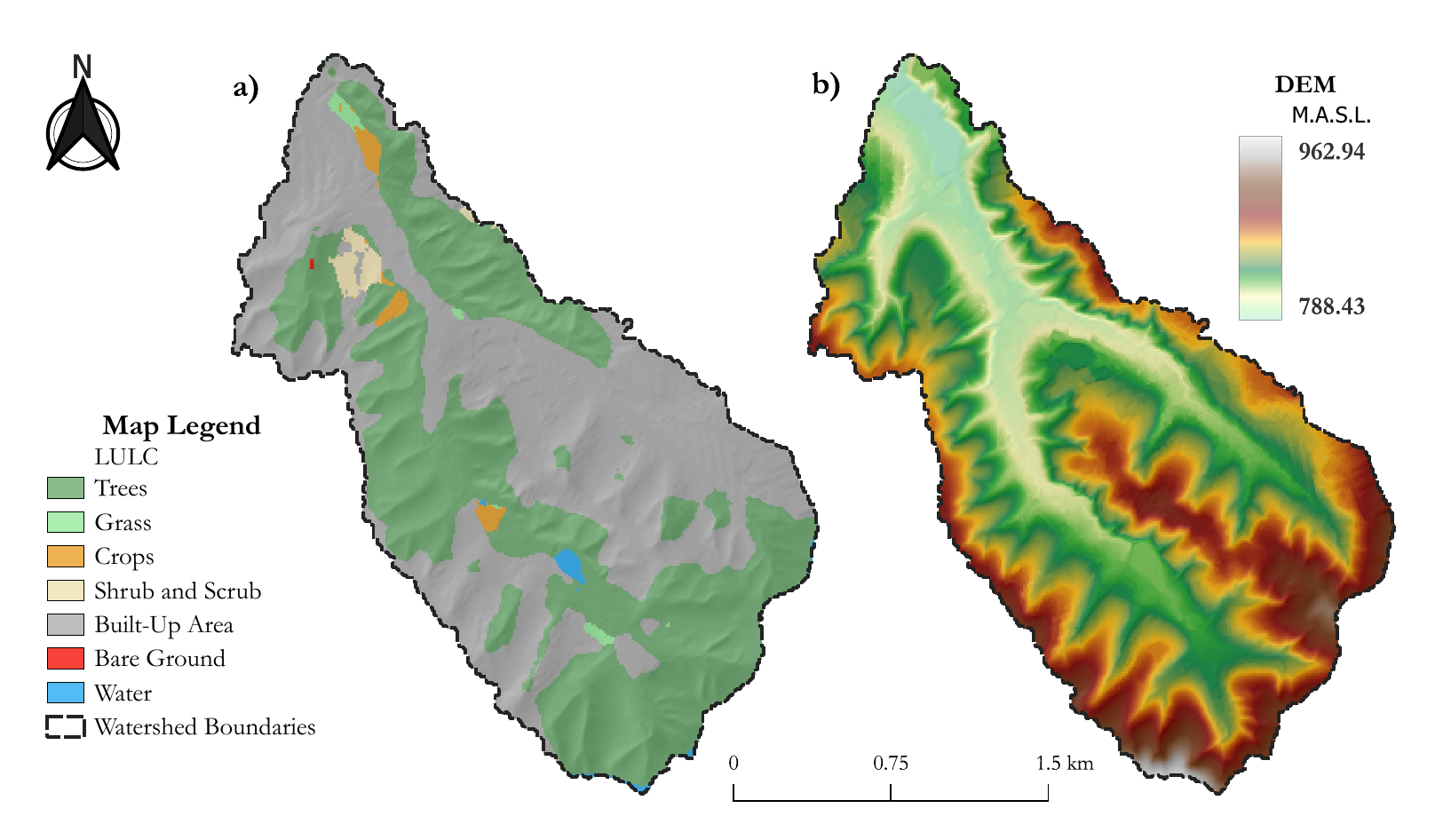}
    \caption{Land use and land cover map (a), hypsometric map of the watershed (b). Elevation was filled and resampled to $\mathrm{50~m}$. Following this, the D-8 watershed algorithm was run, and the watershed boundary was created. All other maps were derived by clipping available rasters into this resulting polygon.}
    \label{fig:BHARI_def}
\end{figure}

The following numerical case studies investigate the system's performance if the orifice and the spillway were controlled through active valves and gates. The digital elevation model collected from airborne LiDAR surveys available on the GeoSampa portal \citep{GeoSampa} and the land use and land cover data were retrieved from the Dynamic World database, which classifies the entire world into eight main classes \citep{brown2022dynamic}. This information is presented in Fig.~\ref{fig:BHARI_def}. To treat the DEM and enhance flow continuity and pathways, a GIS pre-processing in the raw DEM data is performed. Using the topotoolbox \citep{schwanghart2014topotoolbox} one can fill sinks, smooth streams, and impose minimum slopes using functions $\mathtt{fillsinks}$, $\mathtt{klargestconncomps}$, and $\mathtt{imposemin}$ \citep{schwanghart2014topotoolbox}. The watershed is mostly described by the headwaters of the Aricanduva River and, therefore, has a relatively steep average slope of approximately 20\%, indicating that gravitational effects probably govern the hydrodynamics of the catchment. Additionally, this catchment does not have the influence of other upstream reservoirs that would play a role in water storage within the catchment.



\subsubsection{Watershed Properties}
The soil Green-Ampt of suction head, saturated hydraulic conductivity, effective moisture content, and initially stored depth were estimated as 31.5 mm, 2.54 $\mr{cm \cdot h^{-1}}$, 0.476, and 5 mm, respectively based on literature data \citep{rossman2016storm}. The Manning’s roughness coefficient values were estimated for each LULC of Fig.~\ref{fig:BHARI_def}, resulting in 0.015, 0.06, 0.03, 0.12, 0.03, 0.05, and 0.016 $\mr{s \cdot m^{-1/3}}$ for Water, Trees, Grass, Crops, Shrub/Scrub, and Built Areas, respectively \citep{downer2006gridded}.

\subsubsection{Reservoir Stage-Area-Volume}
The reservoir has the area varying with the water surface depth and is described by points of depth x area, presented as follows: 

\begin{equation}
  A(h^{\mr r}(k)) =
    \begin{cases}
      2,833.33h^{\mr r}(k) + 50 & \text{if $h^{\mr r}(k) \leq 0.9$}\\
      2,600 + 59,900(h^{\mr r}(k) - 0.9) & \text{if $0.9 \leq h^{\mr r}(k) \leq 1.9$}\\
      62,500 + 2,080(h^{\mr r}(k) - 1.9) & \text{if $1.9 \leq h^{\mr r}(k) \leq 4.4$} \\
      67,700 + 2,080(h^{\mr r}(k) - 4.4) & \text{if $4.4 \leq h^{\mr r}(k) \leq 6.9,$}
    \end{cases}       
\end{equation}
where the area values are given in m\textsuperscript{2} and the depth values in m.

The depth-varying volume is calculated from the stage-area relationship by integrating this function in Matlab. A linear variation of the area with respect to depth is assumed between two known points of depth-area, allowing the analytical determination of the function that describes the area between two known points.

\subsubsection{Outlet Devices}
The reservoir has a $\mathrm{1~m\times1~m}$ orifice located at the bottom and a spillway at a depth of $\mathrm{4.4~m}$ with 9 m of crest length. A discharge coefficient of 0.61 is assumed for the orifice, resulting in $k_{\mr o} = 5.4039$. If the spillway was uncontrolled (i.e., $q_{\mr s}(k) = k_{\mr s} (h(k)-p)^{1.5})$, the spillway coefficient would result in $k_{\mr f} = 18.9$ (see \citet{gomes2022flood}). However, if the spillway is retrofitted with a controllable device that vertically changes the effective spillway area by a movable vertical gate, the orifice equation discharging at the atmosphere equation can be used, such that:

\begin{equation} \label{equ:gate_equation}
    q_{\mr s}(h) = c_d \overbrace{ \Bigl(u_{\mr s}(k) l_{\mr {ef}} \Bigl (h(k) - p \Bigr ) \Bigr)}^{a_{\mr {ef}}(k)} \sqrt{2g\Bigl (h(k) - p\Bigr )},
\end{equation}
where $u_{\mr s}(k)$ is the control operation between 0 and 1, where 0 represents a fully closed gate, and 1 represents a gate open a depth equals the effective depth in the gate $(h(k) - p)$. Equation \eqref{equ:gate_equation} turns out in the format $q_{\mr s} = k_{\mr s} (h-p)^{\alpha_s}$ as follows:

\begin{equation}
    q_{\mr s}(h) = \overbrace{c_d l_{\mr {ef}} \sqrt{2g}}^{k_{\mr s}} \Bigl (h(k) - p \Bigr )^{3/2},
\end{equation}
with $\alpha_{\mr s} = 3/2$.

Therefore, if one models the spillway as a controllable device using an effective width of 9~m, it results in $k_{\mr s} = 27$ and $\alpha_{\mr s} = 3/2$ for a $c_{\mr d} = 0.68$.

Although controlling the spillway may be desirable, a higher chance of overtopping can occur by reducing the spillway capacity during flood propagation. In cases where the predicted inflow supersedes the reservoir capacity volume, the model assumes the overtopping volume as the spillway outflow volume and emits an alert in the model that this condition is occurring. Typically, this case would have a poor optimization function value by tunning the weights for exceeding $q_{\mr{max}}^{**}$ relatively high. This condition is more likely to occur in reservoirs that are poorly designed.

\begin{figure}
    \centering
    \includegraphics[scale = 0.63]{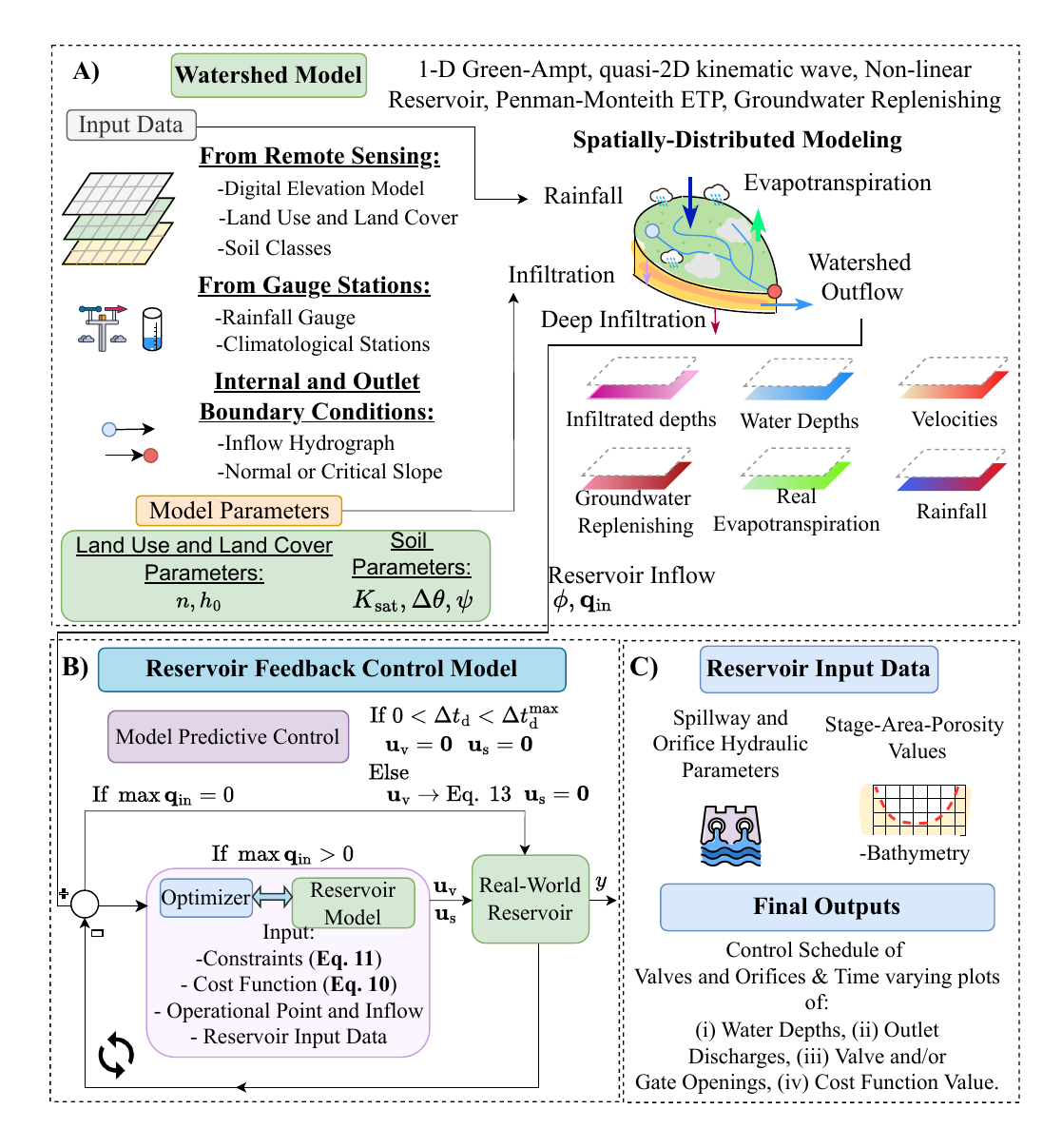}
    \caption{Summary of the modeling aspects of RTC-Stormwater. Part (A) is the watershed model with raster input data representing the terrain's topography, cover, and soils. In addition, spatialized climatologic inputs are considered to represent rainfall and input climatologic data to estimate potential evapotranspiration. Inflow and outlet boundary conditions can be parameterized. The autonomous watershed model has time-varying rasters of infiltrated surface water, rainfall, evapotranspiration depths, and flow velocities. Time-varying vectors of stage and discharge are also outputs of the model, and the latter are the input data for the Reservoir Feedback Control Model (B). The MPC feedback control is presented in Part (B), and the MPC is run considering the inflow hydrograph from the watershed, the cost function and constraints, the current operational points of the system, and the reservoir input data, which are shown in Part (C). Finally, the control schedule of the valve and gates to minimize the flood cost function, as well as time-varying plots of the states and outputs, are shown at the end of the simulation .}
    \label{fig:Methodology}
\end{figure}

A summary of the RTC-Stormwater model is presented in Fig.~\ref{fig:Methodology}.

\subsection{Model Validation - Comparing Kinematic Quasi-2D model with Full Momentum Model}
The RTC-SM model is applied and compared to HEC-RAS 2D in the study area catchment. The hydrographs generated at the catchment outlet are used as the metric to assess model performance. In this scenario, infiltration and initial abstraction are neglected, and only the assessment of the overland flow routing is considered. Moreover, to test only the capacity of the model to predict flood depths, a constant Manning's roughness coefficient of $0.02~\mathrm{s \cdot m^{-1/3}}$ is assumed.

The kinematic wave model is compared with the most advanced solver of HEC-RAS 6.3 - the Shallow-Water-Equation set solved with the Eulerian Method (SWE-EM Stricter Momentum) \citep{Brunner}. The approach is tested by modeling an unsteady-state rainfall of a 100-year storm with the Alternated Blocks hyetograph. 

To make sure the reservoir storage effects do not affect the hydrographs results, an inter-catchment in HEC-RAS is delineated to represent the contributing storage areas that drain to the entry of the reservoir, as shown in Fig.~\ref{fig:BHARI}. Therefore, the effective drainage area was reduced from 4.70~km\textsuperscript{2} to 4.49~km\textsuperscript{2}.

In this scenario, all tested scenarios were simulated with a constant time step that varied according to each model to guarantee numerical stability. The hydrographs of the simulation are compared, and the peak and volume errors are calculated. The models were tested with spatial resolutions of 10 and 30 m.

\subsection{Scenario 1 - Simulating the Water Quantity and Quality Control Under Dynamic Rainfall Events} \label{sec:num_case_2}

This scenario modeled the catchment with a cell size spatial resolution of 10 m. A 44-hour event is tested consisting of two design storms of 10-years, 2-h duration, spanned 6-h each other temporally distributed with the Alternated Blocks Method. Both design storms have 77 mm of rainfall volume each and are temporally distributed using the Alternated Blocks method. This design event shows how the MPC strategy would work under relatively large events occurring sequentially, forcing the MPC strategy to perform under critical forecast.

The MPC approach is compared with the passive scenario, which has valves always open. For the MPC algorithm, a control interval of 1-h, a control horizon of 2-h, and a prediction horizon of 12-h are considered. Furthermore, the MPC optimization function \eqref{equ:optimization_problem} is solved with the $\mathtt{fmincon}$ solver with 120 maximum function evaluations per each initial random initial guess of the solution. Therefore, assuming 5 initial points, the objective function is evaluated 600 times per control horizon, and the solution with a smaller cost function is chosen as the near-optimal control schedule of the prediction horizon. The objective function is evaluated by solving the reservoir dynamics for the prediction horizon and collecting the states to allow for a timely evaluation of the objective function. Therefore, it is important to have a relatively fast model.

The weights of Eq.~\eqref{equ:optimization_problem} are assumed to represent typical detention pond goals. The detention time ($\Delta t_{\mr d}$) is assumed to be 18 hours to represent 1.5 prediction horizons of 12-h and be a relatively sufficient time to sediment solids over the bottom of the detention pond. This parameter can also be adapted to local requirement conditions. In addition, even though the desired detention time is 18 hours, during inter-flood events, the water quality focus might switch to flood control focus, not allowing the total desired detention time.  

To measure the efficiency of the strategy, the values of the objective function given by \eqref{equ:optimization_problem} are plotted, and the minor and major flood times are computed. These are defined as the duration of the reservoir outflow greater than $q_{\mathrm{max}}^*$ and $q_{\mathrm{max}}^{**}$, respectively.

\subsection{Scenario 2 - Simulating the Benefits of RTC under 1-year of Continuous Simulation with Spatial Rainfall and ETP}

This numerical case fully implements the RTC strategy that couples a watershed model and a stormwater reservoir model predictive controller with spatially varied climatologic forcing. The hydrologic and hydrodynamic processes in the watershed with point-source climatologic data are simulated, interpolating the variables to all cells in the domain. This scenario is an example of implementing the model developed in a real-world scenario and for a continuous simulation.

Records of 10-min interval rainfall from 7 rain gauge stations and daily climatologic inputs from 3 meteorological gauge stations are available in the study area and collected for the model inputs, as shown in Fig.~\ref{fig:BHARI}c). The data are then interpolated, and results are obtained so that each cell of the watershed domain has known interpolated rainfall and climatologic inputs. For reservoir control strategies, static approaches of valve full opened (100\%) with partial valve openings of 75\%, 50\%, and 25\%, in addition to the MPC control are compared.

\subsubsection{Climatologic Inputs in the Watershed Model}

The inputs for the quasi 2-D hydrodynamic model can be a combination of distributed or lumped assumptions. Only rainfall and evapotranspiration are assumed as the climatologic inputs; however, other cases and more complex watersheds would require the implementation of inflow hydrographs as internal boundary conditions \citep{Brunner}. An Inverse-Distance-Weightning interpolation (IDW) is performed in the values of each station to estimate spatial values of rainfall and evapotranspiration over time. This procedure is fully detailed in the SI.

\section{Results and Discussion} \label{sec:results}

\subsection{Model Validation}
\begin{figure}
    \centering
    \includegraphics[scale = 0.27]{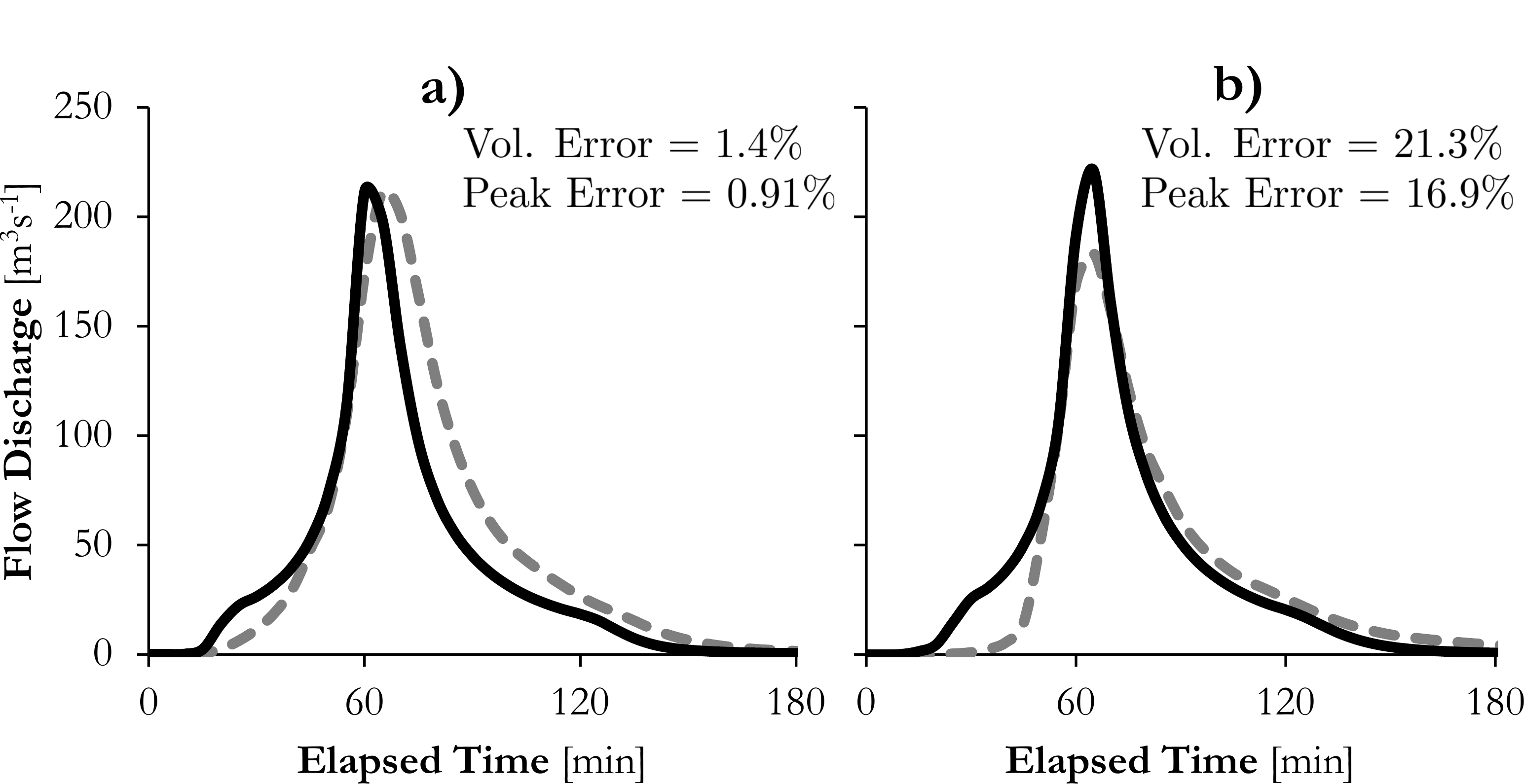}
    \caption{Comparison within the Kinematic-wave solution of the SWE with the full momentum solution. Black lines are the solution of HEC-RAS 2D, and grey dashed lines are the RTC-Kinematic Wave solution. Part a) are results using a 10-m DEM resolution, and b) are results with a 30-m resolution.}
    \label{fig:hydrographs_kinematic}
\end{figure}

The results of comparing RTC-Kinematic 2D with HEC-RAS 2D full momentum under 100-yr storm temporally distributed with the Alternated Blocks method is presented in Fig.~\ref{fig:hydrographs_kinematic}. An unsteady-state rainfall of 100-yr of return period often produces an event in which all terms of the SWE are required, such as local and convective acceleration \citep{akan2021open}. Comparing a relatively simpler kinematic-wave model with the full-momentum unsteady state will probably present a scenario with a larger discrepancy between both models and hence, this analysis is used to assess the model performance to predict the outlet hydrographs. The results presented in Fig.~\ref{fig:hydrographs_kinematic} show a relatively small error between both models, with both models having a great visual agreement, especially for 10-m resolution.  

\subsection{Scenario 1}
\begin{figure}
    \centering
    \includegraphics[scale = 0.65]{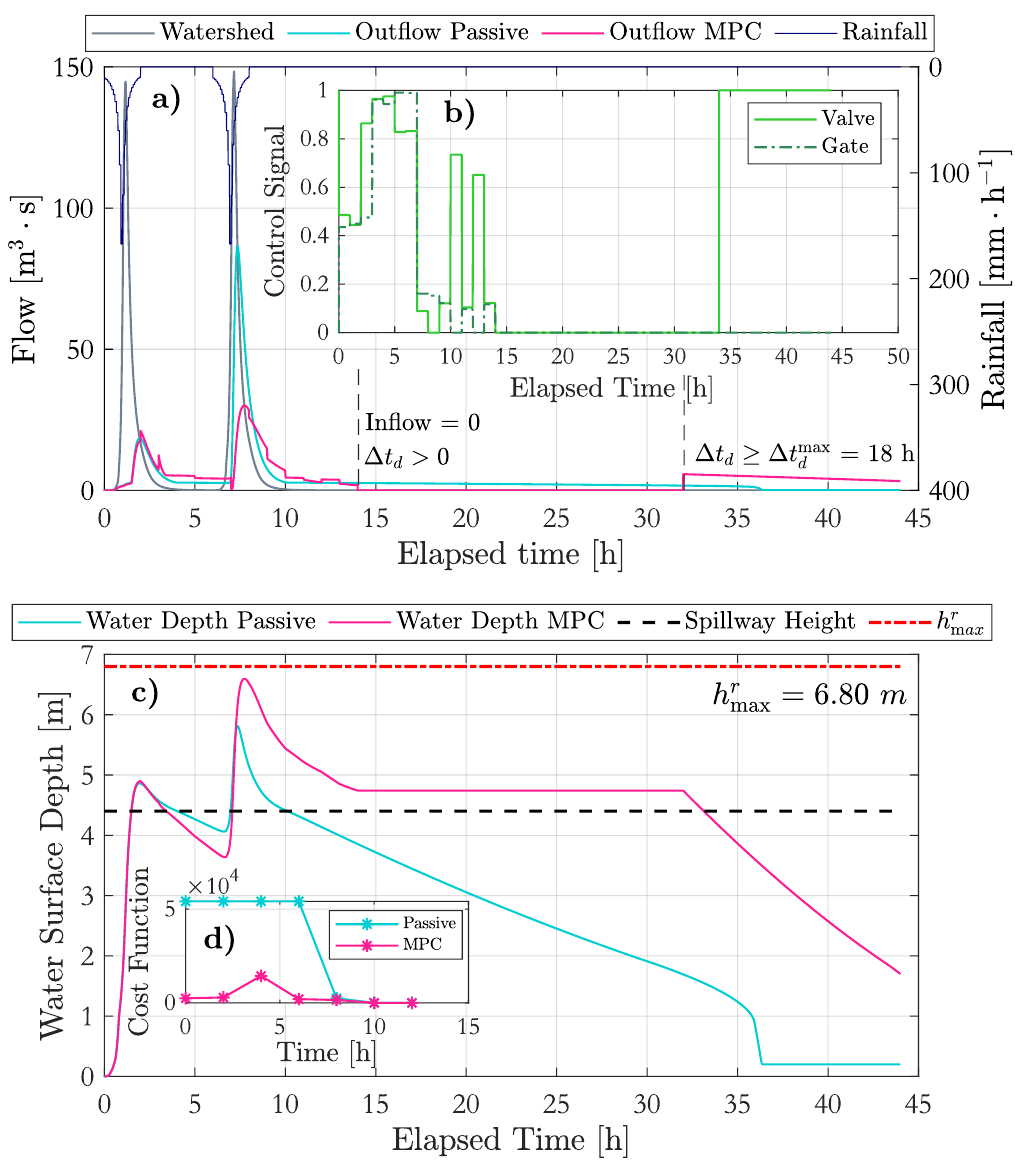}
    \caption{Results of the simulation presented in Scenario 1. Part a) is the discharges from the watershed and the outflows from the reservoir for the passive (i.e., gate and valve fully opened) and active scenario (i.e., MPC controlling valve and gate control schedule). Part b) shows the control schedule of orifice and gate opening derived from minimizing Eq.~\eqref{equ:optimization_problem}, where the control signal values represent the ratio between the controllable flow area and the available flow area. Part c) shows stage hydrographs in the reservoir for passive and active scenarios, and Part d) shows the objective function values of Eq.~\eqref{equ:optimization_problem} for each prediction horizon. Variable $\Delta t_{\mr d}$ is the detention time, and $\Delta t_{\mathrm{max}}$ is the required detention time.}
    \label{fig:hydrographs_design_storms}
\end{figure}

The modeling results of the Scenario 2 are presented in Fig.~\ref{fig:hydrographs_design_storms}. Part a) shows the inflow and outflow hydrographs for passive and active MPC-controlled scenarios. The inflow hydrographs had peaks of approximately $\mathrm{148~m^3 \cdot s^{-1}}$, resulting from two consecutive rainfall events of 2-h and 10 years of recurrence interval. Although the passive scenario shows a great peak flow reduction for the first storm, the second peak flow was approximately $\mathrm{80~m^3 \cdot s^{-1} }$ compared to $\mathrm{30~m^3 \cdot s^{-1}}$ resulting from the MPC approach. The passive scenario provided 41\% of peak flow reduction, while the MPC approach provided 79\% attenuation. 

The MPC approach scenario shows that reservoir outflows are smaller than $\mathrm{40~\mathrm{m^3 \cdot s^{-1}}}$, that is, smaller than the defined threshold for large flood events $q_{\mr{max}}^{**}$ (see Fig.~\ref{fig:schematics}). The MPC approach predicts the future inflow hydrograph using a 12-hour time span, implements controls for each hour, and moves 2-h in advance, receiving another forecast of the inflow hydrograph up to the total simulation time. Therefore, the MPC controller could predict the second and larger flow wave before the first inflow hydrograph peak. 

Although releasing flows greater than $q_{\mr{max}}^*$ would increase the costs associated with $\rho_{\mr{q,k}}$ from \eqref{equ:optimization_problem}, by releasing flows at a rate smaller than $q_{\mr{max}}^{**}$, the method avoids a larger penalization from $\rho_{**}$. The MPC algorithm decides to release more flows after the first storm to have available volume and depth to control discharges and flood depths to the desired levels actively. One can see in Fig.~\ref{fig:hydrographs_design_storms}b) that the active control decides to open the valves and gates during the inter-event duration to allow more future inflows to be stored in the reservoir and to be able to temporally partially close the gates and valves during the second inflow hydrograph to have a better flow mitigation. One can also see in Fig.~\ref{fig:hydrographs_design_storms}c) that the MPC maintains the flow for approximately 18-h and releases it afterward following Eq.~\eqref{equ:control_detention}. The cost functions for each scenario for each control horizon are presented in Fig.~\ref{fig:hydrographs_design_storms}d), with optimization results of MPC significantly outperforming the passive scenario.


\subsection{Scenario 2}
\begin{figure}
    \centering
    \includegraphics[scale = 0.6]{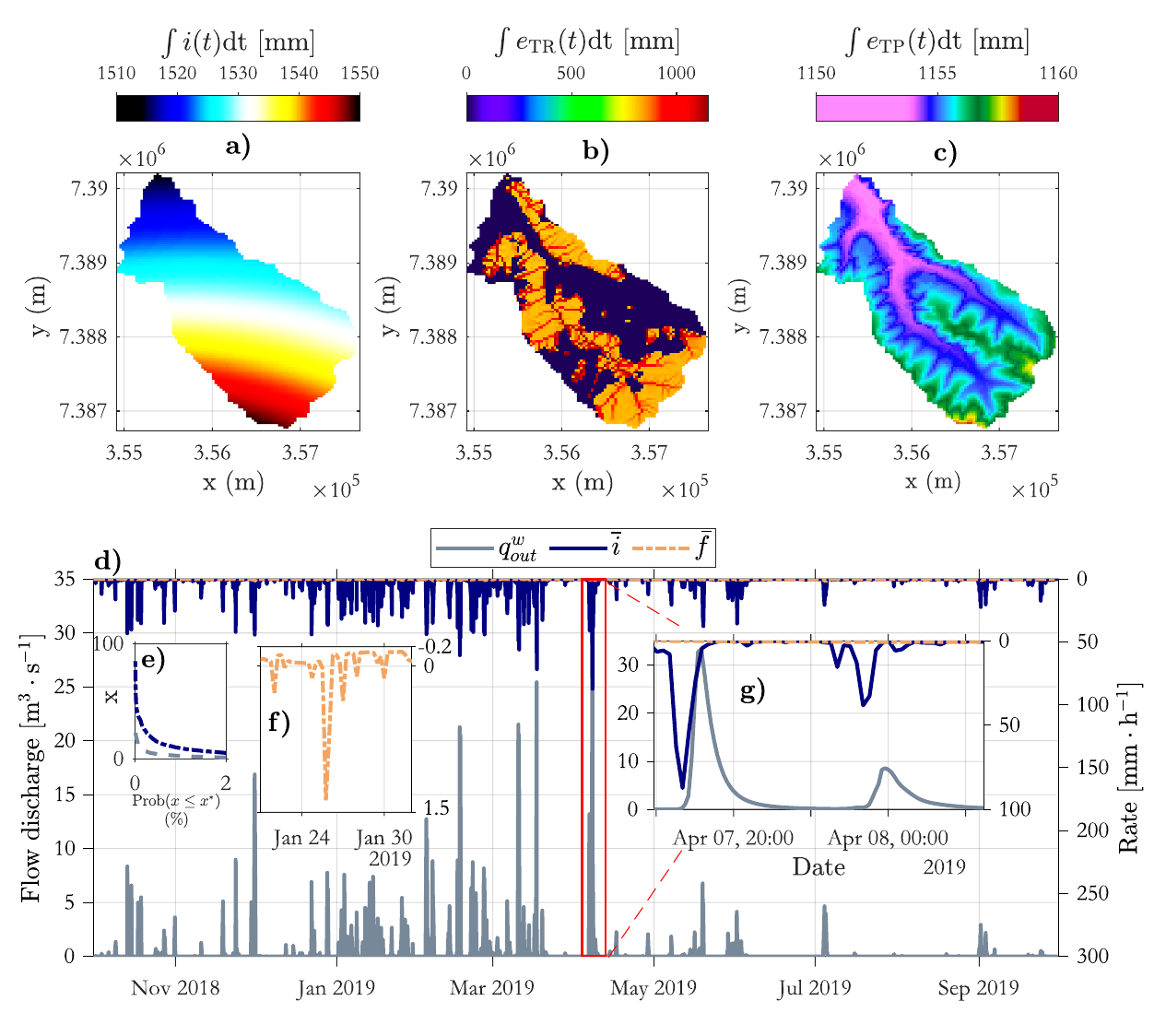}
    \caption{Rainfall ETP and ETR for the hydrologic year of 2018-2019 in the Aricanduva Catchment, Sao Paulo. Part (a), (b), and (c) are the 1-yr cumulative rainfall real evapotranspiration, and potential evapotranspiration, respectively. Part (d) is the aerial average hyetograph, infiltration, and outlet discharge, where (e) shows the duration curve of discharge and rainfall, (f) shows a detail of the surface aerial net flux ($\bar{f}$), and (g) details the largest event recorded in the year.}
    \label{fig:rain_etp_etr}
\end{figure}

A summary of the watershed results is illustrated in Fig.~\ref{fig:rain_etp_etr}. The cumulative values of rainfall, real evapotranspiration, and potential evapotranspiration are shown in parts (a)-(c) of this figure.  A 1-year rainfall volume of approximately 1500 mm; a potential evapotranspiration of nearly 1150 mm and a real evapotranspiration of approximately 1000 mm are observed. Some areas had higher real evapotranspiration values due to water availability in the soil. 

Fig.~\ref{fig:rain_etp_etr} part d) shows the 1-yr hyetograph and hydrograph, with details of the event of April 7 highlighted in part g) and a duration curve chart of rainfall and discharge shown in e). To illustrate the surface aerial average net flux ($\bar{f}$), an insert chart is presented in f). The surface net flux considers the balance between infiltration rate and evapotranspiration rate, and details of this calculation are presented in the SI. The model's hydrologic-hydrodynamic watershed modeling component can be used to understand spatialized hydrologic information such as those presented in Fig.~\ref{fig:rain_etp_etr}. 

Using the 1-year watershed outlet hydrograph shown in Fig.~\ref{fig:rain_etp_etr}, the MPC algorithm is set to optimize the valve and gate control while controlling the detention time when no inflows are predicted. The MPC routine had a computational time of approximately 18 h to optimize the 4,380 control horizons of 2-h for a 1 year of inflow hydrograph. The objective function was evaluated at least 2 million times, showing the importance of a fast-cost function. 

\begin{figure}
    \centering
    \includegraphics[scale = 0.67]{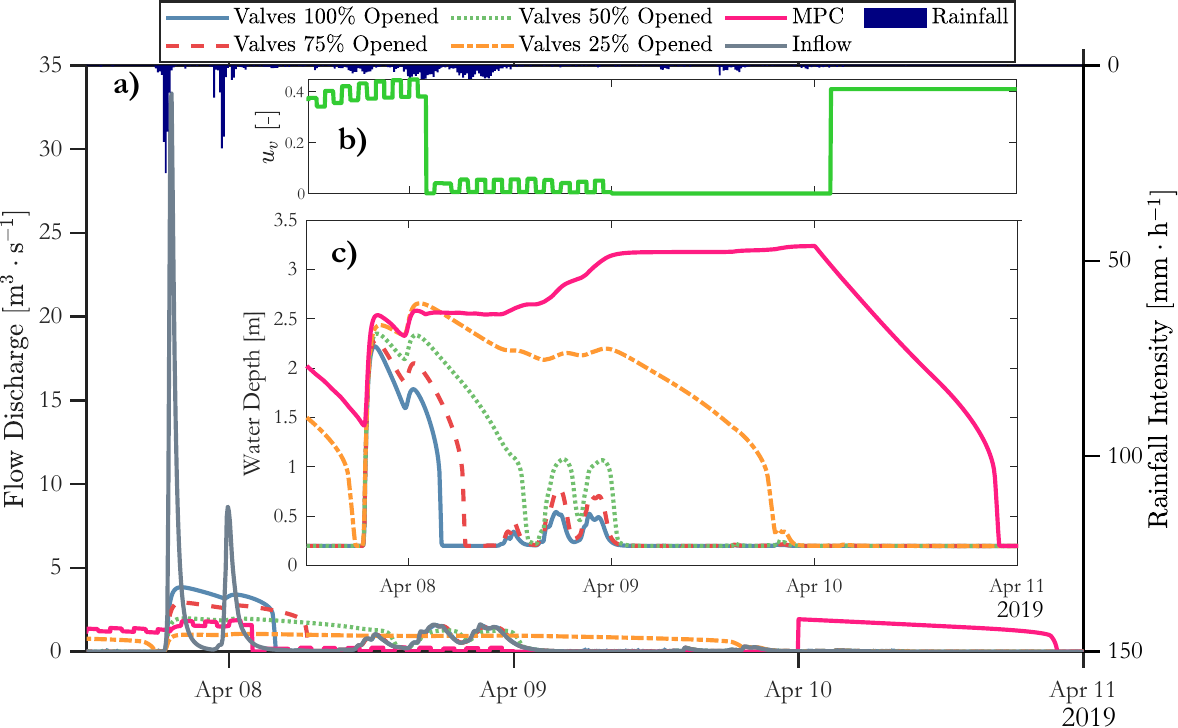}
    \caption{Example of a simple performance evaluation for a rainfall event using static rules of valve partially open with valve openness of 100, 75, 50, and 25\% compared to the MPC controlled valve. Part (a) are the reservoir outlet hydrographs and the inflow hydrograph, while (b) is the control schedule of the MPC approach for this event, and (c) is the stage hydrograph in the reservoir.}
    \label{fig:hydrographs_peak}
\end{figure}

The detailed event of April 7 (see Fig.~\ref{fig:rain_etp_etr}g)) is expanded up to 11 April to show the performance of the MPC in comparison with other passive strategies, as shown with the outlet hydrographs for the scenarios with valves 100, 75, 50, and 25\% opened in contrast to the MPC control in Fig.~\ref{fig:hydrographs_peak}. Part (a) shows the inflow and outflow hydrographs, with all cases with peak flow reductions larger than 25~$\mr{m^3 \cdot s^{-1}}$. Most of the controls, however, failed to mitigate the peak flows after the first two peaks. That means that the static operation of the reservoir did not have a peak flow mitigation effect for minor storms. On the other hand, the MPC control kept the water up to the expected detention time and then released it using the threshold for releasing the stored volume ($q_{\mr t}^{\mr *}$). The valve operation and water depth are shown in Fig.~\ref{fig:hydrographs_peak}b) and Fig.~\ref{fig:hydrographs_peak}c), respectively. These results imply that retrofitted reservoirs with active controlling capacity (i.e., time-varying controlling devices) can mitigate large and minor storms. 

In Fig.~\ref{fig:hydrographs_peak}c), the stage hydrographs in the reservoir are depicted. It is noted that the MPC approach held more water inside the reservoir longer after reaching the first two peaks since the optimization problem changed from flood to water quality control. Additionally, a valve opening larger than 25\% seems to have minor mitigation effects for the minor upcoming floods. 

\begin{figure}
    \centering
    \includegraphics[scale = 0.71]{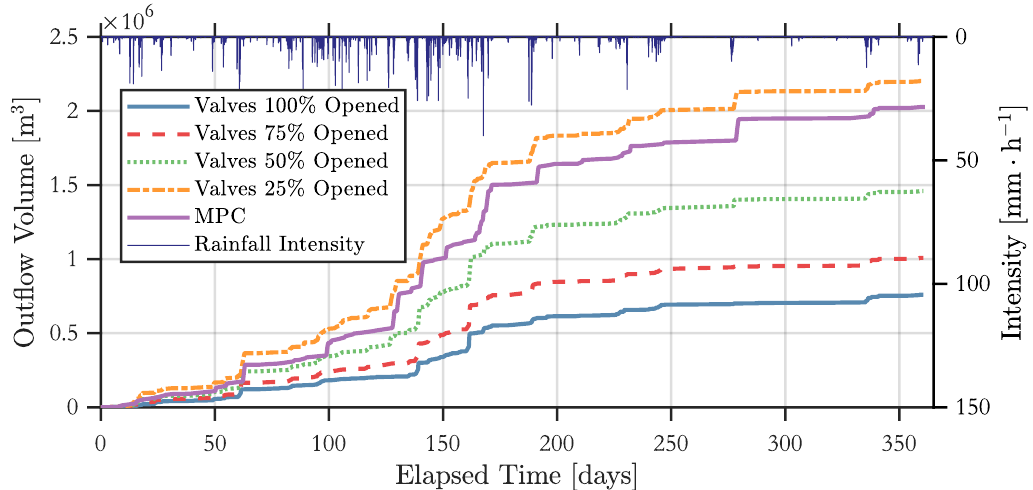}
    \caption{Treated volume for varied types of static controls and the MPC control.}
    \label{fig:treated_volumes}
\end{figure}

\begin{figure}
    \centering
    \includegraphics[scale = 0.75]{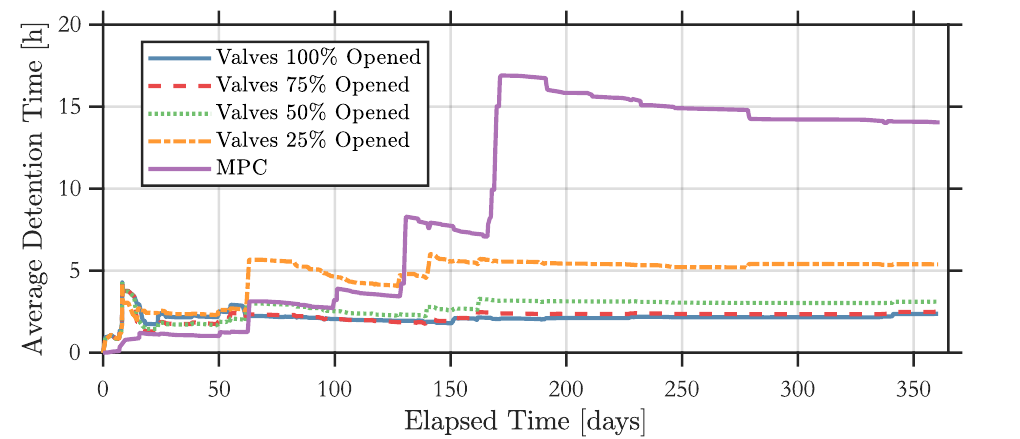}
    \caption{Average detention time curves for 1-yr of continuous simulation, where the MPC alternative aims to provide 18-h of detention time.}
    \label{fig:detention_times}
\end{figure}

\begin{figure}
    \centering
    \includegraphics[scale = 0.65]{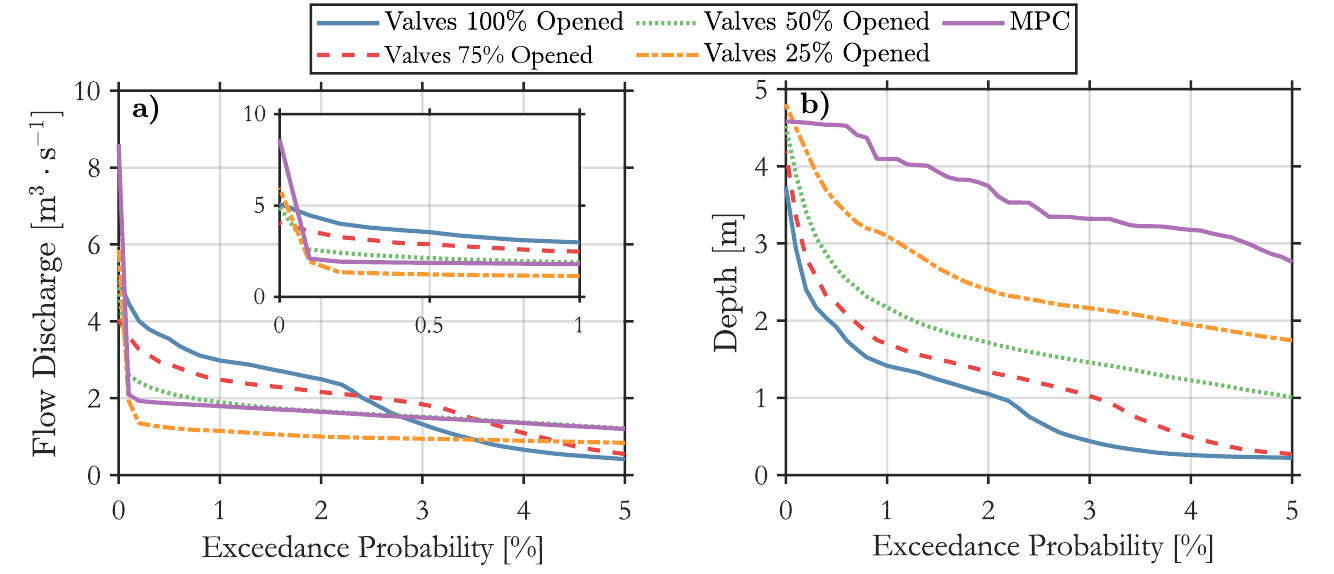}
    \caption{Duration curves for 1-yr of continuous simulation, where (a) is flow duration curve for scenarios of 100, 75, 50, and 25\% in the valve openness, compared to the MPC controlled case Part (b) represents the stage duration curve.}
    \label{fig:duration_curves}
\end{figure}

Fig.~\ref{fig:treated_volumes} shows the evolution of the treated volume, which is only calculated when no inflow hydrographs are predicted on the control horizon and when the water depth is larger than a minimum threshold. More volume is treated using a valve opening of 25\% compared to MPC; however, as shown in Fig.~\ref{fig:detention_times}, the MPC approach has a longer detention time, which is a proxy representation of the water quality state of the system. The duration curves of the flow discharges and depths are shown in Fig.~\ref{fig:duration_curves}, indicating the probability of reaching a flow or depth larger than a threshold. It is observed from Fig.~\ref{fig:duration_curves}a) that the passive scenario has a sharp change in the flow duration curve after 2\%. The stage duration curve on Fig.~\ref{fig:duration_curves}b) shows great variability in the probabilities of exceeding certain depths for the control strategies, with the MPC scenario presenting larger depths than the other controls.

\section{Conclusions} \label{sec:conclusions}
A watershed-distributed continuous hydrodynamic model coupled with a reservoir model is integrated, and a model predictive control algorithm is used optimally to control the valves and gates of a real-world reservoir. The reservoir control adjusts its objectives based on predicted inflows. By switching the focus from flood mitigation to runoff detention, it was possible to achieve flood and proxy water quality control represented by specific minimum detention times. The general conclusions of this paper are described as follows:

\begin{itemize}
    \item Reservoir hydraulic devices typically designed for large storm events can be adapted to mitigate recurrent small ones if retrofitted with real-time MPC. The full outflow capacity enables the conveyance of large flood events as expected; however, in recurrent events, no flow mitigation would be achieved with a passive control strategy since valves and gates are always open and inflows are smaller than the flow capacity. If reservoirs are retrofitted with the proposed MPC, the operation of valves and gates can achieve desirable flow mitigation performance by actively controlling the outflow capacity to meet the flow mitigation requirements, as illustrated in the scenarios tested. This conclusion is important for this paper's case study since several reservoirs on the path of the Aricanduva River have reports of poor mitigation effects for recurrent floods.

\item The MPC strategy presented in this paper allows for balancing flood mitigation and water quality by shifting the focus of the control approach based on the predicted hydrograph. Two flood severity discharges are defined to change the weights of the optimization function according to the predicted inflow hydrographs. These discharges are tuned and can represent flow thresholds that local regulation constraints can have (e.g., minor flood threshold, major flood threshold). If no floods are predicted, then the model switches to focus on maximizing the detention time up to the desired maximum detention time defined. The results of this control approach showed superior performance for flood mitigation under critical design storms and 1-year of continuous simulation compared to passive scenarios. For the continuous simulation, the volume treated for the MPC approach is similar to the 25\% opened valves scenario that had 6 hours of average detention time compared to the nearly 14 hours detention time provided by the MPC approach. A similar runoff volume is treated for both cases. This result indicates that not only a superior performance of flood mitigation is achieved with MPC, but also a better proxy water quality is obtained since a larger average detention time is achieved. 

\item More research is required to investigate the trade-offs among the MPC control variables of the prediction horizon, control horizon, and control steps. Although fixed and assumed in this paper, these parameters change the efficiency of the runoff control of reservoirs.
\end{itemize}

In addition to the general conclusions, the model validation, scenario 1, and scenario 2 specific conclusions are:

\begin{itemize}
    \item \textit{Model Validation}: The watershed quasi 2D kinematic wave hydrodynamic model provided similar results when compared to the HEC-RAS 2D full momentum solver, with a better performance when using 10-m resolution than 30-m spatial resolution.
    \item \textit{Scenario 1}: During two consecutive 2-hour, 10-year storms that occurred 6 hours apart, the MPC algorithm effectively managed the valve and spillway of the detention pond. This not only improved flood mitigation performance but also ensured that the stored volume had the desired detention time, as compared to the passive scenario where the valves and spillways were fully open.
    \item \textit{Scenario 2}: A 1-yr continuous simulation results indicate that the MPC can regulate the flow discharges at the cost of generally maintaining a typical larger water level stored in the reservoir compared to passive scenarios. The MPC approach outperformed all passive scenarios regarding flood mitigation and increased detention times to the desired level.
\end{itemize}

The results presented show how a detention pond can perform better in flood mitigation and water quality treatment by changing the operation of the reservoir from a passive gravity control approach to an active and real-time opening and closing of valves and gates. Sources of uncertainty, such as (i) rainfall forecasting, (ii) conceptual model simplifications, and (iii) lack of correcting states with techniques such as Kalman filters, are important aspects that could be incorporated into future studies. 

Future studies must address some important limitations to understand the robustness of MPC applied to real-world stormwater reservoirs. First, the uncertainty in the rainfall predictions must be assessed to understand whether the MPC can still perform better than a passive scenario under these circumstances. In addition, the hydrological model itself has its own uncertainty in the conceptualization that can also underestimate the inflow hydrographs. Therefore, a self-controlled system that can update states and model parameters with observations is desired. The hydrological models used to simulate the watershed processes provide simple estimates of the system states that can be further corrected by real-time field measurements, autocorrecting, and auto-calibrating the model with a Kalman Filter approach. Investigating the aforementioned issues is a future direction to improve understanding of how MPC applied to detention ponds can improve flood mitigation and water quality treatment. Extensive testing of the model against other models and with field observations is warranted. 

In addition, the proposed control strategy can also be subjected to failures if extremely large storms are rapidly temporally distributed within one prediction horizon. For instance, if there is a large hurricane or a rapidly occurring flash flood within the predicted time horizon, and the reservoir is currently storing water for water quality purposes, the system might not be able to handle the event because of its limited storage capacity. Therefore, the numerical approach presented here can also be used solely to investigate the effects of flood control or water quality control instead of both combined.

To successfully adapt the approach presented here to other reservoirs, modelers have to tune the weights of the objective functions representing the trade-offs between the flow mitigation thresholds for large and minor flood events, the expected detention time, and ultimately, the expected maximum flow discharge released after reaching the water quality detention time. These parameters can be easily adapted to local regulations and estimated with hydrological data, allowing for an improvement in the performance of detention ponds for flood mitigation and runoff detention.

\section{Data Availability Statement}
Some or all data, models, or code generated or used during the
study are available in a repository or online in accordance with
funder data retention policies. All software can be freely downloaded in \citet{RTC_Code}.

\section*{Acknowledgment}
The present study was supported by CAPES Ph.D Scholarship and by the San Antonio River Authority. Prof. Taha acknowledges the support from the National Science Foundation through Grant Number 2151392.

\section*{Supplemental Materials}
Supplementary data related to this article can be found at \url{https://github.com/marcusnobrega-eng/RTC---Flood-and-Water-Quality}. 

\printcredits

\bibliographystyle{cas-model2-names}
\bibliography{cas-refs}



\end{document}